\documentclass[11pt,a4paper,reqno]{amsart}

\usepackage[latin1]{inputenc}
\usepackage[english]{babel}
\usepackage{amsmath}
\usepackage[numbers]{natbib}
\usepackage{graphicx}
\usepackage{amsmath,bm}
\usepackage{booktabs}
\usepackage{graphicx}
\usepackage{multirow}
\usepackage{setspace}
\usepackage{braket}
\usepackage[hmargin=3cm,vmargin={3cm,4cm}]{geometry}

\usepackage{etoolbox}
\newcommand*{\affaddr}[1]{#1} 
\newcommand*{\affmark}[1][*]{\textsuperscript{#1}}

\begin{document}

\title{Multiscale modelling of \textit{de novo} anaerobic granulation}

\author[A.~Tenore et al.]{A.~Tenore\protect\affmark[1] \and F.~Russo\affmark[1] \and M.R.~Mattei\affmark[1] \and B.~D'Acunto\affmark[1] \and G.~Collins\affmark[2] \and L.~Frunzo\affmark[1]}

 \maketitle

 {\footnotesize
  \begin{center}
\affaddr{\affmark[1]Department of Mathematics and Applications, University of Naples "Federico II", via Cintia 1, Montesantangelo, 80126, Naples, Italy}\\
\affaddr{\affmark[2]Microbial Communities Laboratory, School of Natural Sciences, National University of Ireland, Galway, University Road, H91 TK33, Ireland}\\
 \end{center}}

{\footnotesize
\begin{center}
Corresponding authors: A.~Tenore, \texttt{alberto.tenore@unina.it}; M.R.~Mattei, \texttt{mariarosaria.mattei@unina.it}
 \end{center}}

\begin{abstract}

A multiscale mathematical model is presented to describe the \textit{de novo} granulation and the evolution of multispecies granular biofilms within a continuous reactor. The granule is modelled as a spherical free boundary domain with radial symmetry. The equation which governs the free boundary is derived from global mass balance considerations and takes into account the growth of sessile biomass and the exchange fluxes with the bulk liquid. Starting from a vanishing initial value, the expansion of the free boundary is initiated by the attachment process, which depends on the microbial species concentrations within the bulk liquid and their specific attachment velocity. Non-linear hyperbolic PDEs model the growth of the sessile microbial species, while quasi-linear parabolic PDEs govern the dynamics of substrates and invading species within the granular biofilm. Nonlinear ODEs govern the evolution of soluble substrates and planktonic biomass within the bulk liquid. The model is applied to an anaerobic granular-based system and solved numerically to test its qualitative behaviour and explore the main aspects of \textit{de novo} anaerobic granulation: ecology, biomass distribution, relative abundance, dimensional evolution of the granules and soluble substrates and planktonic biomass dynamics within the reactor. The numerical results confirm that the model accurately describes the ecology and the concentrically-layered structure of anaerobic granules observed experimentally, and is able to predict the effects of some significant factors, such as influent wastewater composition, granulation properties of planktonic biomass, biomass density and hydrodynamic and shear stress conditions, on the process performance.
\end{abstract}


\maketitle

\section{Introduction}
\label{n1}

Biofilms are complex, dense and compact aggregates composed by microbial cells immobilized in a self-produced matrix of extracellular polymeric substances (EPS) \cite{flemming2010biofilm}. Several trophic groups coexist in such structures and deeply interact with each other through synergistic and antagonistic activities. Although natural biofilms typically develop as planar layers attached to suitable surfaces, under specific conditions the aggregation occurs due to the self-immobilization of cells into approximately spherical-shaped granules \cite{trego2019novo}. The process leading to the formation of these aggregates is known as granulation. In particular, the term \textit{de novo} granulation is used when the process is initiated by individual microbial cells and flocs, in opposition to the case in which the granulation process takes place starting from an inoculum in granular form.

In recent years, granular systems have become increasingly popular in the field of wastewater treatment. Compared to the suspended biomass systems, the denser, stronger and more regular structure of the biofilm granules leads to better settling properties \cite{liu2002essential,liu2004state} which allow higher biomass concentrations \cite{baeten2019modelling} and reduced-footprint reactors \cite{liu2002essential}. Furthermore, in contrast to other biofilm systems where the biomass develops on solid supports, granular systems are based on spherical and constantly moving microbial aggregates. The movement and shape mitigate boundary layer resistances and enhance the mass transfer of substrates across the biofilm granule \cite{baeten2019modelling}. For these reasons, granular biofilms have been successfully developed in different reactor configurations, for various processes, such as aerobic, anaerobic and partial nitritation-anammox treatments \cite{trego2019granular}.

The main drawback of granular-based systems is represented by the start-up phase, due to the complexity of the mechanisms and phenomena which contribute to the success of the granulation process \cite{li2015significant}. In the last years, many researchers have studied and explored this process and numerous theories have been proposed. Hydrodynamic conditions generated by the liquid up-flow velocity, the gas production, the particle-particle collision, the mixing systems and the reactor geometry are universally recognized as the key factors for the granulation process and the entire life cycle of the granules \cite{liu2002essential,trego2019granular,lettinga1980use}. Indeed, suitable hydrodynamic conditions are needed to initiate the granulation process by promoting and improving the aggregation of planktonic biomass \cite{lemaire2008micro} and, at the same time, avoiding its complete washout. Moreover, intense hydrodynamic conditions induce high shear forces on the granule surface which influence dimension, shape, structure and density of the granules \cite{liu2002essential,di2006influence} and regulate a continuous process of aggregation and breaking which leads to the formation of an increasing number of granules. High shear forces are thought to stimulate the production of EPS which represents a further beneficial factor for the granulation as it increases cell surface hydrophobicity \cite{trego2019granular}. Several studies also consider the granulation process as the result of an organized process driven by pioneering microbial species which have specific characteristics \cite{pol2004anaerobic,wiegant1988spaghetti}. 

In \cite{pol2004anaerobic}, various theories concerning the anaerobic sludge granulation are collected and described. The most widespread theory asserts that anaerobic granulation process is favoured by key microorganisms such as \textit{Methanosaeta} \cite{trego2019granular,pol2004anaerobic,jian1993study,fang2000microbial}. Such acetoclastic methanogens have filamentous structures and good adhering properties and initiate the granulation process by forming a central nucleus and favouring the immobilization of other methanogens and synergistically functioning bacterial groups \cite{macleod1990layered,zheng2006monitoring}. In this context, various studies \cite{li2015significant,zhang2012acyl,li2014characterization} report that quorum sensing plays an essential role by regulating the transition of some \textit{Methanosaeta} species from short to filamentous cells. In the initial phase, the nucleus presents a filamentous appearance and achieves a spherical shape due to the rolling effect of the hydraulic shear forces \cite{trego2019granular,pol2004anaerobic}. In a second phase, the nucleus develops into a granule and acetogens and acidogens attach on its surface and grow syntrophically with acetoclastic methanogens \cite{trego2019granular,pol2004anaerobic,macleod1990layered,vanderhaegen1992acidogenesis}. The result is a concentrically-layered structure with an archaeal core constituted by \textit{Methanosaeta}. This theory is supported by several experimental evidences showing layered structures in anaerobic granules \cite{sekiguchi1999fluorescence,batstone2004influence,collins2005distribution}. In any case, the granulation process still remains not fully understood and further studies are required to optimize the efficiencies of this process.

In this framework, mathematical modelling represents a valuable tool to describe, explore and study the granulation process, the life cycle of the biofilm granules and the performances of granular-based systems. In the last years, the modelling of granular biofilm systems has stimulated a growing interest due to the relevance of this topic in engineering, biological and industrial fields. Numerous models have been proposed to mainly describe aerobic \cite{de2007kinetic}, anaerobic \cite{batstone2004influence,odriozola2016modeling,feldman2017modelling,doloman2017modeling} and anammox \cite{volcke2010effect,volcke2012granule,vangsgaard2012sensitivity} processes involved in such systems. A first classification can be introduced according to the approach used: continuum models simulate the evolution of the granular biofilm in a quantitative and deterministic way, while discrete models, such as individual-based \cite{doloman2017modeling,picioreanu2005multidimensional,xavier2007multi} and cellular automata models \cite{skiadas2002new}, are able to represent the multidimensional structural heterogeneity of granular biofilm but provide results including elements of randomness and introduce stochastic effects into the solutions \cite{mattei2018continuum}. Most models of granular biofilms \cite{batstone2004influence,odriozola2016modeling,feldman2017modelling,volcke2010effect,volcke2012granule} are based on the continuum approach introduced in \cite{wanner1986multispecies} for one-dimensional planar biofilms, and model the granule as a spherical free boundary domain, which evolves as a result of the microbial metabolic processes and the mass exchange with the surrounding environment. Among them, most describe the dynamic evolution of the granule fixing the final steady-state dimension \cite{batstone2004influence,feldman2017modelling,volcke2010effect,volcke2012granule}. 

In any case, some significant aspects of the granular biofilm growth are not considered exhaustively by the models present in the literature. According to \cite{baeten2019modelling}, only two models \cite{batstone2004influence,seok2003integrated} consider the attachment process, which plays a key role in the formation and evolution of granular biofilms. No model takes into account the invasion process: the colonization of a pre-existing biofilm mediated by motile planktonic cells living in the surrounding environment, which can penetrate the porous matrix of the biofilm and convert to sessile biomass. Moreover, all continuum models fix a non-zero initial size of the domain and thus describe an initial configuration where the granular biofilm is already formed. Finally, according to the exclusion principle presented in \cite{klapper2011exclusion}, all biofilm models based on the approach introduced in \cite{wanner1986multispecies} lead to restrictions on ecological structure. 

Concerning the aim and the numerical studies pursued in literature, most works focus on system performances by describing the biological purification process and the removal of soluble substrates from wastewater. Someone focuses on the biofilm granule, paying attention to the dimensional evolution \cite{odriozola2016modeling} and to the distribution of sessile biomass within the biofilm at the steady-state \cite{odriozola2016modeling,feldman2017modelling,volcke2010effect,volcke2012granule,vangsgaard2012sensitivity}. However, no continuous model fully describes the \textit{de novo} granulation process by considering the initial formation and ecology of the biofilm granule. Only the individual-based model introduced in \cite{doloman2017modeling} focuses on the \textit{de novo} formation of anaerobic granules based on a discrete approach.

In this work, we propose a multiscale model which describes the \textit{de novo} granulation process, and incorporates the mesoscopic granular biofilm processes within a continuous granular-based reactor. For this purpose, following the approach proposed in \cite{mavsic2012persistence,mavsic2014modeling} in the case of one-dimensional planar biofilms, the presented model couples the macroscopic reactor mass balances with the mesoscopic granular biofilm model here derived by using a continuum approach \cite{wanner1986multispecies}. The model accounts for the growth of both granular attached and planktonic biomasses, and includes the main microbial exchange processes involved, such as attachment, detachment and invasion. The \textit{de novo} granulation process is modelled by assuming that all biomass initially present in the reactor is in planktonic form. Mathematically, this corresponds to consider a vanishing initial value of the granule radius which represents the free boundary under the assumption of radial symmetry. The biofilm formation is initiated by the attachment process, which leads to consider a space-like free boundary. This mathematical problem has been discussed in \cite{d2019free} and is applied here for the first time to model the genesis of granular biofilms. The granule formation and expansion are governed by the following processes: microbial growth, attachment, invasion and detachment. Attachment initiates the life of biofilms and is regarded as the complex phenomenon whereby pioneering microbial cells in planktonic form attach to a surface and develop in the form of sessile aggregate \cite{palmer2007bacterial}. However, as reported above, the formation of a biofilm granule is the result of the interaction and aggregation of microbial cells and flocs without the involvement of a surface. Therefore, in this work, attachment is accounted as the flux of microbial mass which aggregates, switches its phenotype from planktonic to sessile and initiates the granulation. It is modelled as a linear function of the concentrations of the planktonic species, each of them characterized by a specific attachment velocity. The invasion process is included for the first time in the modelling of granular biofilms by extending the mathematical formulation proposed in \cite{d2015modeling} for one-dimensional planar biofilms to a spherical domain. This allows to remove the restrictions on ecological structure highlighted in \cite{klapper2011exclusion}. Furthermore, the bulk liquid is modelled as a perfectly mixed medium where soluble substrates and planktonic biomass are found and influenced by the operational parameters of the reactor, the microbial metabolic activities and the processes of mass exchange with the biofilm. The mathematical model has been derived for a generic granular-based reactor and applied to the anaerobic granulation process in order to test the model behaviour and study the genesis, evolution and ecology of anaerobic granules. Different numerical studies have been carried out to investigate how the granulation properties of the planktonic biomass, the biomass density of the granules, the shear forces and the composition of the influent wastewater can affect the evolution of the process. The results include the dimensional evolution and ecology of the granule, in terms of biomass distribution and relative abundance, and the variation of soluble substrates and planktonic biomass within the reactor. 
 
The paper is organized as follows. In Section \ref{n2}, the derivation of the model is carried out by presenting all assumptions, variables, equations, initial and boundary conditions. Section \ref{n3} describes the biological case to which the model is applied. Numerical studies are reported and discussed in detail in Section \ref{n4} and \ref{n5}, respectively. Finally, the conclusions of the work and the future goals are outlined in Section \ref{n6}.

\section{Mathematical Model} 
\label{n2}

In this work, the granular biofilm reactor is modelled as a completely mixed continuous system where $N_G$ identical biofilm granules are immersed. As shown in Fig. \ref{2.1}, two different scales are considered in the model: the bioreactor macroscale and the granule mesoscale. Three components are considered within the granular biofilm: the sessile biomass which constitutes the solid matrix, the planktonic biomass which is found in the channels and voids and the soluble substrates dissolved in the liquid phase. Meanwhile, planktonic biomass and soluble substrates are considered within the bulk liquid of the reactor. These components interact and influence each other as a result of biological, physical and chemical processes. In the following, the modelling of both the granule and reactor scales is discussed, introducing all the processes, assumptions, variables, equations, initial and boundary conditions involved.

\begin{figure*}
 \fbox{\includegraphics[width=1\textwidth, keepaspectratio]{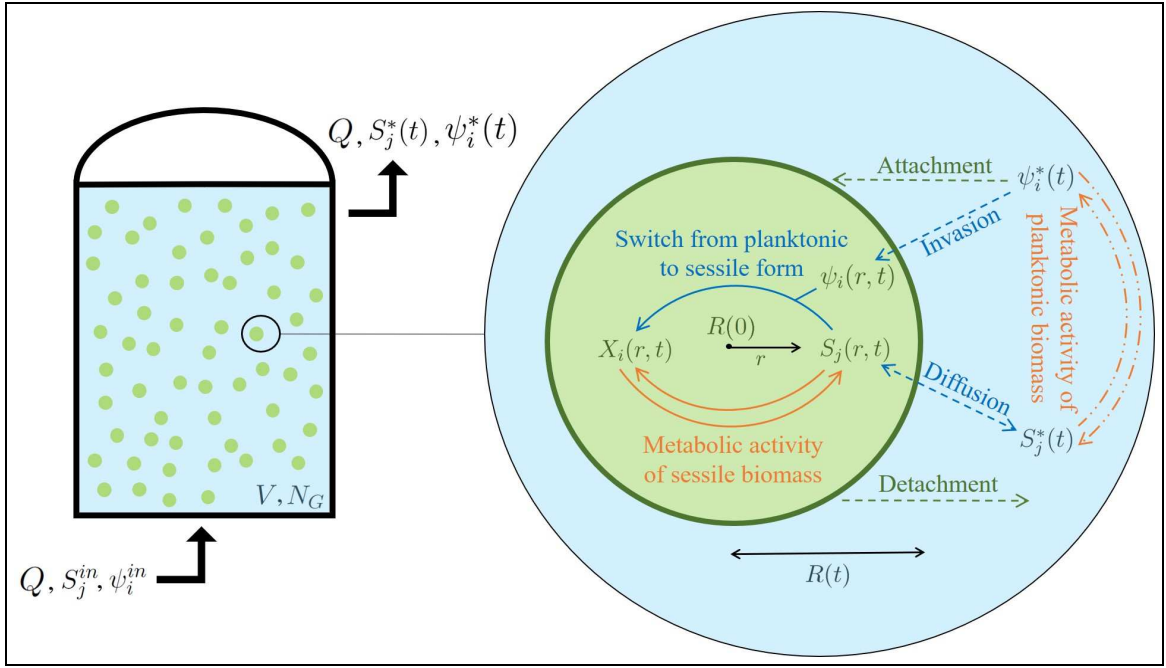}}   
 \caption{Multiscale representation of the model. The bioreactor is modelled as a perfectly mixed continuous system (on the left), having volume $V$, where $N_G$ biofilm granules are immersed. A focus on a single granule is presented on the right, with all processes considered in the model. The granule has a zero initial radius $R(0)$, which varies over time due to the effect of various biological processes. Metabolic processes within the granule are carried out by the sessile biomass $X_i(r,t)$, which grows by converting the substrates dissolved in the biofilm liquid $S_j(r,t)$, while metabolic processes within the bulk liquid are carried out by the planktonic biomasses $\psi^*_i(t)$, which grows by converting the substrates dissolved in the bulk liquid $S^*_j(t)$. The superficial exchange processes of attachment and detachment are considered at the interface granule-bulk liquid. Moreover, invasion processes are modelled: the planktonic biomass $\psi_i(r,t)$ invades the solid matrix of the granule and switches its phenotype from planktonic to sessile. Finally, process of diffusion of substrates across the granule is included in the model. Solid arrows: processes within the granule. Dash-dot arrows: processes within the bulk liquid. Dash arrows: exchange processes between granule and bulk liquid.}
          \label{f2.1} 
 \end{figure*}

 \subsection{Modelling granule scale} 
 \label{n2.1} 

Under the assumption of radial symmetry, the biofilm granule is modelled as a spherical free boundary domain whose spatial evolution is completely described by the evolution of the radius $R(t)$. A vanishing initial value $R(0)=0$ is considered to model the initial granulation. All variables involved in the biofilm modelling are considered as functions of time $t$ and space $r$, where $r$ denotes the radial coordinate. Consequently, the granule centre is located at $r = 0$.

The model takes into account the dynamics of three components, expressed in terms of concentration: $n$ microbial species in sessile form $X_i(r,t)$; $n$ microbial species in planktonic form $\psi_i(r,t)$; $m$ dissolved substrates $S_j(r,t)$.

The volume occupied by planktonic cells is considered negligible due to the small particle size. The density of the granule $\rho$ is assumed to be constant and equal for all microbial species. By dividing sessile species concentration $X_i$ by $\rho$, biomass volume fractions $f_i(r,t)$ are achieved. $f_i$ are constrained to add up to unity \cite{rahman2015mixed}. In summary, the model components which describe the granular biofilm compartment are:

\begin{equation}      \label{2.1}
X_i,\  i=1,...,n,\ {\bf X}=(X_1,...,X_n),
\end{equation}	

\begin{equation}      \label{2.2}
f_i=\frac{X_i}{\rho},\  i=1,...,n,\ {\bf f}=(f_1,...,f_n),
\end{equation}	

\begin{equation}      \label{2.3}
\psi_i,\  i=1,...,n,\ \mbox{\boldmath $\psi$}=(\psi_1,...,\psi_n),
\end{equation}	

\begin{equation}      \label{2.4}
S_j,\  j=1,...,m,\ {\bf S}=(S_1,...,S_m).
\end{equation}	

Based on the continuum approach introduced in \cite{wanner1986multispecies} for one-dimensional planar biofilms, a system of partial differential equations (PDEs) in a spherical free boundary domain is derived from mass balance considerations, under the assumption of radial symmetry. Hyperbolic PDEs model the distribution and growth of sessile biomass $f_i(r,t)$ and parabolic PDEs describe the diffusion and conversion of soluble substrates $S_j(r,t)$. Further parabolic PDEs govern the process of invasion and conversion of planktonic cells $\psi_i(r,t)$. 

Based on the aggregation properties of the planktonic biomass living in the bulk liquid, attachment phenomena govern the initial granulation process, while the further biofilm evolution is significantly affected by detachment phenomena. Attachment and detachment contribute to the microbial mass exchange occurring between granules and bulk liquid and are included in the model as continuous and deterministic processes. Finally, as introduced in \cite{d2015modeling} in the case of one-dimensional planar biofilm, the invasion process is considered to describe the phenomena of granule colonization by planktonic cells. Such cells penetrate the porous matrix of the biofilm from the surrounding medium and contribute to the development of the biofilm.

Under the assumption of radial symmetry, the mass balance set up for a generic component in a differential volume of the spherical domain, leads to the following equation:

\begin{equation}                                        \label{2.5}
			\frac{\partial c(r,t)}{\partial t} +\frac{1}{r^2} \frac{\partial}{\partial r}(r^2 J_r(r,t))
			 =r_c(r,t), 
\end{equation}	

where $c(r,t)$ is the concentration of a generic component in the spherical domain, $J_r$ is the advective and/or diffusive flux in the radial direction and $r_c(r,t)$ is the transformation term.

The transport of sessile biomass is modelled as an advective process. Hence, by expressing the advective flux of the $i^{th}$ sessile microbial species in the radial direction as 

\begin{equation}                                        \label{2.6}
J_{r,i}(r,t)=u(r,t) X_i(r,t),
\end{equation}	

where $u(r,t)$ is the biomass velocity, Eq.\eqref{2.5} takes the following form:

  \[	 
			\frac{\partial X_i(r,t)}{\partial t} +\frac{1}{r^2} \frac{\partial}{\partial r}(r^2 u(r,t) X_i(r,t))
			 = \rho r_{M,i}(r,t,{\bf X},{\bf S}) + \rho r_i(r,t,\mbox{\boldmath $\psi$},{\bf S}),
  \]
\begin{equation}                                        \label{2.7}
i=1,...,n, 0 \leq r \leq R(t),\ t>0,
\end{equation}	

where $r_{M,i}^{}(r,t,{\bf X},{\bf S})$ and $r_i(r,t,\mbox{\boldmath $\psi$},{\bf S})$ are the specific growth rates due to sessile and planktonic species, respectively.

By dividing Eq.\eqref{2.7} by $\rho$ and by considering that $f_i=\frac{X_i}{\rho}$ yields

  \[
			\frac{\partial f_i(r,t)}{\partial t} +\frac{1}{r^2} \frac{\partial}{\partial r}(r^2 u(r,t) f_i(r,t))
			 =r_{M,i}(r,t,{\bf f},{\bf S})+r_i(r,t,\mbox{\boldmath $\psi$},{\bf S}),
  \]	 
\begin{equation}                                        \label{2.8}
i=1,...,n, 0 \leq r \leq R(t),\ t>0,
\end{equation}	

    \[	 
			\frac{\partial f_i(r,t)}{\partial t} + f_i(r,t) \frac{\partial u(r,t)}{\partial r} + \frac{2 u(r,t) f_i(r,t)}{r} + u(r,t) \frac{\partial f_i(r,t)}{\partial r}
			 =r_{M,i}(r,t,{\bf f},{\bf S})+
  \]
\begin{equation}                                        \label{2.9}
  +r_i(r,t,\mbox{\boldmath $\psi$},{\bf S}), \ i=1,...,n, 0 \leq r \leq R(t),\ t>0.
\end{equation}	

Summing Eq.\eqref{2.9} over all sessile microbial species $i$ and considering that $\sum_{i=1}^{n}f_i=1$, it follows:

		\begin{equation}                                        \label{2.10}		
	     \frac{\partial u(r,t)}{\partial r} = -\frac{2 u(r,t)}{r} + G(r,t,\textbf{f},\textbf{S}, \mbox{\boldmath $\psi$}, 0 < r \leq R(t),\ t>0,
	\end{equation}
	
	where $G(r,t,\textbf{f},\textbf{S}, \mbox{\boldmath $\psi$})=\sum_{i=1}^{n}(r_{M,i}(r,t,\textbf{f},\textbf{S})+r_i(r,t,\mbox{\boldmath $\psi$},\textbf{S}))$. This differential equation governs the evolution of the biomass velocity $u(r,t)$. 

By imposing the flux of the $i^{th}$ sessile microbial species equal to $0$ at $r=0$, it follows from Eq.\eqref{2.6} that $u(0,t)=0$. Considering this result and integrating Eq.\eqref{2.10}, the integral expression of $u(r,t)$ is achieved:

 \begin{equation}                                       \label{2.11}
   u(r,t)=\frac{1}{r^2}\int_{0}^{r} r'^2 G(r',t,\textbf{f},\textbf{S}, \mbox{\boldmath $\psi$}) dr', 0 < r \leq R(t),\ t>0.
 \end{equation}
	
Substituting Eq.\eqref{2.10} into Eq.\eqref{2.9} yields

   \[	 
\frac{\partial f_i(r,t)}{\partial t} +  u(r,t)\frac{\partial f_i(r,t)}{\partial r}
			 =r_{M,i}(r,t,{\bf f},{\bf S})+r_i(r,t,\mbox{\boldmath $\psi$},{\bf S}) - f_i(r,t) G(r,t,\textbf{f},\textbf{S}, \mbox{\boldmath $\psi$}),
  \]
\begin{equation}                                        \label{2.12}
		 i=1,...,n, 0 \leq r \leq R(t),\ t>0.
\end{equation}	

Eq.\eqref{2.12} describes the transport and growth of the sessile microbial species $i$ across the granular biofilm under the assumption of radial symmetry.

Compared to the equation reported in \cite{wanner1986multispecies} for a planar biofilm, Eq.\eqref{2.12} presents a different expression of $u(r,t)$ and the additional reaction term $r_i(r,t,\mbox{\boldmath $\psi$},{\bf S})$ due to the invasion phenomenon.

Eq.\eqref{2.5} can be applied to soluble substrates and planktonic species. In these cases, the transport of planktonic biomass and soluble substrates is modelled as a diffusive flux and expressed as

\begin{equation}                                        \label{2.13}
J_{r,\psi_i}(r,t)=-D_{\psi,i}\frac{\partial \psi_i(r,t)}{\partial r},
\end{equation}
	
and 

\begin{equation}                                        \label{2.14}
J_{r,j}(r,t)=-D_{S,j}\frac{\partial S_j(r,t)}{\partial r},
\end{equation}

where $D_{\psi,i}$ and $D_{S,j}$ denote the diffusivity coefficient of the planktonic species $i$ and the soluble substrate $j$ in the biofilm, respectively.

Then, parabolic diffusion-reaction PDEs are derived from Eq.\eqref{2.5}:

   \[	 
	    \frac{\partial \psi_i(r,t)}{\partial t}-D_{\psi,i}\frac{\partial^2 \psi_i(r,t)}{\partial r^2} - \frac{2 D_{\psi,i}}{r} \frac{\partial \psi_i(r,t)}{\partial r}=
   r_{\psi,i}(r,t,\mbox{\boldmath $\psi$},{\bf S}),
  \]
 \begin{equation}                                        \label{2.15}
 \ i=1,...,n, 0 < r < R(t),\ t>0,
 \end{equation}
 
   \[	 
  \frac{\partial S_j(r,t)}{\partial t}-D_{S,j}\frac{\partial^2 S_j(r,t)}{\partial r^2} - \frac{2 D_{S,j}}{r} \frac{\partial S_j(r,t)}{\partial r}=
   r_{S,j}(r,t,{\bf f},{\bf S}),
  \]
 \begin{equation}                                        \label{2.16}
	   \ j=1,...,m, 0 < r < R(t),\ t>0,
 \end{equation}

where $r_{\psi,i}(r,t,\mbox{\boldmath $\psi$},{\bf S})$ is the conversion rate of planktonic species $i$ and $r_{S,j}(r,t,{\bf f},{\bf S})$ is the conversion rate of soluble substrate $j$.

The free boundary evolution is described by the variation of the radius $R(t)$ over time. This is affected by microbial growth and processes of attachment and detachment occurring at the surface of the biofilm. In particular, the attachment flux of the $i^{th}$ planktonic species is supposed to be linearly dependent on the concentration of the planktonic species $i$ in the bulk liquid $\psi^*_i(t)$ and is expressed as:

  \begin{equation}                                        \label{2.17}
    \sigma_{a,i}(t)=\frac{v_{a,i}\psi^*_i(t)}{\rho},\ i=1,...,n, 
 \end{equation}

where $v_{a,i}$ is the attachment velocity of the planktonic species $i$.

By summing Eq.\eqref{2.17} over all planktonic species, the total attachment flux is achieved:

  \begin{equation}                                        \label{2.18}
    \sigma_a(t)=\frac{\sum_{i=1}^{n}v_{a,i}\psi^*_i(t)}{\rho}. 
 \end{equation}
 
The detachment is modelled as a quadratic function of the granule radius $R(t)$ \cite{abbas2012longtime}:

\begin{equation}                                        \label{2.19}
	    \sigma_d(t)=\lambda R^2(t),
\end{equation} 

where $\lambda$ is the detachment coefficient and is supposed to be equal for all microbial species.

The global mass balance on the spherical domain gives:
 
 \begin{equation}                                       \label{2.20}
  \frac{\partial}{\partial t}\int_{0}^{R(t)} 4 \pi r^2 \rho dr
    =\rho A(t) (\sigma_a(t)-\sigma_d(t))+
   \int_{0}^{R(t)} 4 \pi r^2 \rho G(r,t,\textbf{f},\textbf{S}, \mbox{\boldmath $\psi$}) dr,
 \end{equation}
 
 where $A(t)$ is the area of the spherical granule and is equal to $4 \pi R^2(t)$.
 
By dividing Eq.\eqref{2.20} by $4 \pi \rho$ and by considering $u(R(t),r)$ from Eq.\eqref{2.11}, it follows:
 
 \begin{equation}                                       \label{2.21}
  \frac{\partial}{\partial t}\int_{0}^{R(t)} r^2 dr
    =R^2(t) (\sigma_a(t)-\sigma_d(t))+
   \int_{0}^{R(t)} r^2 G(r,t,\textbf{f},\textbf{S}, \mbox{\boldmath $\psi$}) dr,
 \end{equation}
 
  \begin{equation}                                       \label{2.22}
  \frac{1}{3}\frac{\partial R^3(t)}{\partial t}
    =R^2(t) (\sigma_a(t)-\sigma_d(t))+
   R^2(t) u(R(t),t),
 \end{equation}
 
   \begin{equation}                                       \label{2.23}
  \dot R(t)
    = \sigma_a(t)-\sigma_d(t) + u(R(t),t).
 \end{equation}
  
The latter equation governs the time evolution of the free boundary domain. 

The total mass of the sessile community and the mass of the $i^{th}$ sessile microbial species within the granule can be calculated as follows:

 \begin{equation}                                       \label{2.24}
  m_i(t)
    =   \int_{0}^{R(t)} 4 \pi r^2 \rho f_i(r,t) dr,\ i=1,...,n, 
 \end{equation}
  
 \begin{equation}                                       \label{2.25}
  m_{tot}(t) = \sum_{i=1}^{n} m_i(t)
    =   \frac{4}{3} \pi \rho R^3(t).
 \end{equation}

 \begin{table}[!htb]
\begin{tiny}
 \begin{center}
 \begin{spacing}{3}
 \rotatebox{90}{%
 \begin{tabular}{lccc}
\hline
\multirow{2}*{{\textbf{Equations}}} & {\textbf{Initial condition}} & \multicolumn{2}{c}{{\textbf{Boundary condition}}} \\
\cmidrule(l){2-4}
 & {\textbf{$t=0$}} & {\textbf{$r=0$}}  & {\textbf{$r = R(t)$}} \\
 \hline
  
 $\frac{\partial f_i(r,t)}{\partial t} +  u(r,t)\frac{\partial f_i(r,t)}{\partial r} = r_{M,i}(r,t,{\bf f},{\bf S})+r_i(r,t,{\bm \psi},{\bf S})- f_i(r,t) G(r,t,\textbf{f},\textbf{S}, \bm \psi))$  & & & $f_i(R(t),t) = \frac{v_{a,i}\psi^*_i(t)}{\sum_{i=1}^{n}v_{a,i}\psi^*_i(t)}$ \ when $\sigma_{a}-\sigma_{d}>0$ \\

 $\frac{\partial \psi_i(r,t)}{\partial t}-D_{\psi,i}\frac{\partial^2 \psi_i(r,t)}{\partial r^2} - \frac{2 D_{\psi,i}}{r} \frac{\partial \psi_i(r,t)}{\partial r}=
   r_{\psi,i}(r,t,{\bm \psi},{\bf S})$ & & $\frac{\partial \psi_i}{\partial r}(0,t)=0$ & $\psi_i(R(t),t))=\psi^*_i(t)$ \\

$\frac{\partial S_j(r,t)}{\partial t}-D_{S,j}\frac{\partial^2 S_j(r,t)}{\partial r^2} - \frac{2 D_{S,j}}{r} \frac{\partial S_j(r,t)}{\partial r}=r_{S,j}(r,t,{\bf f},{\bf S}))$ & & $\frac{\partial S_j}{\partial r}(0,t)=0$ & $S_j(R(t),t))=S^*_j(t)$ \\

  $\frac{\partial u(r,t)}{\partial r} = -\frac{2 u(r,t)}{r} + G(r,t,\textbf{f},\textbf{S}, \bm \psi)$  & & $u(0,t)=0$ & \\
     
 $\dot R(t)= \sigma_a(t)-\sigma_d(t)+u(R(t),t)$ & $R(0)= 0$ & & \\

  $V \dot \psi^*_i(t)=Q(\psi^{in}_i-\psi^*_i(t))- A(t) N_G D_{\psi,i} \frac{\partial \psi_i(R(t),t)}{\partial r}+r^*_{\psi,i}(t,{\bm \psi^*},{\bf S^*})-\sigma_{a,i}(t) \rho A(t) N_G$  & $\psi^*_i(0)=\psi^*_{i,0}$ & &\\

   $V \dot S^*_j(t)=Q(S^{in}_j-S^*_j(t))- A(t) N_G D_{S,j} \frac{\partial S_j(R(t),t)}{\partial r}+r^*_{S,j}(t,{\bm \psi^*},{\bf S^*})$ & $S^*_j(0)=S^*_{j,0}$ & & \\

  \hline
 \end{tabular}
 }
 \caption{Model equations and initial and boundary conditions} \label{t2.1}
 \end{spacing}
 \end{center}
 \end{tiny}
 \end{table}

\subsection{Modelling reactor scale} 
\label{n2.2} 

As already mentioned, the reactor is modelled as a completely mixed continuous system. Thus, all the quantities referring to the bulk liquid dynamics are equal at every point and are dependent on time. The variables considered in the bulk liquid are $n$ planktonic biomasses and $m$ soluble substrates, both expressed in terms of concentration ($\psi^*_i(t)$ and $S^*_j(t)$, respectively). Such concentrations vary over time due to biological processes, operational parameters of the reactor and mesoscopic granule processes. In summary, the model components which describe the bulk liquid compartment are:

\begin{equation}      \label{2.26}
\psi^*_i,\  i=1,...,n,\ {\bm \psi^*}=(\psi^*_1,...,\psi^*_n),
\end{equation}	

\begin{equation}      \label{2.27}
S^*_j,\  j=1,...,m,\ {\bf S^*}=(S^*_1,...,S^*_m).
\end{equation}	

Accordingly, a system of ordinary differential equations (ODEs) derived from mass balance considerations is considered to describe the dynamics of planktonic biomass and soluble substrates within the bulk liquid:

   \[	 
	   V \dot \psi^*_i(t)=Q(\psi^{in}_i-\psi^*_i(t))- A(t) N_G D_{\psi,i} \frac{\partial \psi_i(R(t),t)}{\partial r}+r^*_{\psi,i}(t,{\bm \psi^*},{\bf S^*})+
  \]	
 \begin{equation}                                       \label{2.28}
-\sigma_{a,i}(t) \rho A(t) N_G,\ i=1,...,n\,\ t>0, 
 \end{equation}
 
    \[	 
	   V \dot S^*_j(t)=Q(S^{in}_j-S^*_j(t))- A(t) N_G D_{S,j} \frac{\partial S_j(R(t),t)}{\partial r} +r^*_{S,j}(t,{\bm \psi^*},{\bf S^*}),
  \]	
 \begin{equation}                                       \label{2.29}
\ j=1,...,m,\,\ t>0.
 \end{equation}
 
where $V$ is the volume of the bulk liquid assumed equal to the reactor volume, $Q$ is the continuous flow rate, $\psi^{in}_i$ is the concentration of the planktonic species $i$ in the influent, $S^{in}_j$ is the concentration of the substrate $j$ in the influent, $r^*_{\psi,i}(t,{\bm \psi^*},{\bf S^*})$ and $r^*_{S,j}({t,\bm \psi^*},{\bf S^*})$ are the conversion rates for $\psi_i^*$ and $S_j^*$, respectively.

Eq.\eqref{2.28} represents the mass balance of the $i^{th}$ microbial species in planktonic form. In particular, the mass variation within the reactor (first member) is due to the continuous mass flow in and out of the reactor (first term of the second member), the exchange flux between the bulk liquid and the granular biofilms (second term of the second member), the growth and decay in the bulk liquid (third term of the second member), the exchange flux related to attachment processes (fourth term of the second member). Notably, no contribution by detachment to planktonic species is considered in Eq.\eqref{2.28}. Indeed, detachment phenomena within granular biofilm-based reactors involve large pieces of granule which detach from the mother granule. Such biofilm pieces can attach to existing granules, constitute new granules or disintegrate and become planktonic biomass. Due to the uncertainty and complexity of such mechanisms, any contribution by detachment to planktonic or attached biomass is neglected and may be considered in future works.

Similarly, Eq.\eqref{2.29} represents the mass balance of the $j^{th}$ soluble substrate. In this case, the mass variation within the reactor (first member) is due to the continuous mass flow in and out of the reactor (first term of the second member), the exchange flux between the bulk liquid and the granular biofilms (second term of the second member) and the consumption and/or production occurring in the bulk liquid and mediated by the planktonic biomass (third term of second member).

\subsection{Initial and boundary conditions} 
\label{n2.3}

The processes involved in a granular biofilm reactor are described by Eqs.\eqref{2.10}, \eqref{2.12}, \eqref{2.15}, \eqref{2.16}, \eqref{2.23}, \eqref{2.28} and \eqref{2.29}. In order to integrate such equations, it is necessary to specify initial and boundary conditions.

As mentioned above, the \textit{de novo} granulation is modelled by considering an initial configuration whereby only planktonic biomass is supposed to be present in the reactor. Hence, a vanishing initial condition is coupled to Eq.\eqref{2.23} which describes the variation of the granule radius over time:

\begin{equation}                                          \label{2.30}
	    R(0)= 0.
\end{equation}

The following initial conditions are considered for Eq.\eqref{2.28} and Eq.\eqref{2.29}:

  \begin{equation}                                        \label{2.31}
	   \psi^*_i(0)=\psi^*_{i,0}, \ i=1,...,n,
 \end{equation}
 
 \begin{equation}                                        \label{2.32}
	   S^*_j(0)=S^*_{j,0},\ j=1,...,m,
 \end{equation}
  
where $\psi^*_{i,0}$ and $S^*_{j,0}$ are the initial concentrations of the $i^{th}$ planktonic species and the $j^{th}$ soluble substrate within the bulk liquid, respectively.
  
Eqs.\eqref{2.10}, \eqref{2.12}, \eqref{2.15} and \eqref{2.16} refer to the biofilm domain and do not require initial conditions, since the extension of the biofilm domain is zero at $t = 0$.

The boundary condition for Eq.\eqref{2.12} at the interface granule-bulk liquid $r = R(t)$ depends on the sign of the mass flux at the interface. When the free boundary is a space-like line ($\sigma_{a}-\sigma_{d}>0$), there is a mass flux from bulk liquid to biofilm, thus the boundary condition depends on the concentration of planktonic biomass in the bulk liquid:

  \begin{equation}                                        \label{2.33}  
   f_i(R(t),t) = \frac{v_{a,i}\psi^*_i(t)}{\sum_{i=1}^{n}v_{a,i}\psi^*_i(t)}, \ i=1,...,n,\ t>0.
  \end{equation} 

Meanwhile, when the free boundary is a time-like line ($\sigma_{a}-\sigma_{d}<0$), the biomass concentration at the interface is regulated exclusively by the internal points of the biofilm domain and the condition \eqref{2.33} is not required.

 
 
For both parabolic systems \eqref{2.15} and \eqref{2.16}, a no-flux condition is fixed at the granule centre $r = 0$, while boundary conditions at the interface biofilm-bulk liquid $r = R(t)$ are related to the solutions of Eq.\eqref{2.28} and Eq.\eqref{2.29}, which represent the concentrations of planktonic species and soluble substrates within the bulk liquid:

 \begin{equation}                                        \label{2.34}
   \frac{\partial \psi_i}{\partial r}(0,t)=0,\ \psi_i(R(t),t))=\psi^*_i(t),\ i=1,...,n,\ t>0,
 \end{equation}
 
 \begin{equation}                                        \label{2.35}
   \frac{\partial S_j}{\partial r}(0,t)=0,\ S_j(R(t),t))=S^*_j(t),\ j=1,...,m,\ t>0.
 \end{equation}

 Finally, as previously mentioned, the boundary condition for Eq.\eqref{2.10} is given by:
 
  \begin{equation}                                        \label{2.36}
   u(0,t)=0,\ t>0.
  \end{equation}
 
In conclusion, the model is based on Eqs.\eqref{2.10}, \eqref{2.12}, \eqref{2.15}, \eqref{2.16}, \eqref{2.23}, \eqref{2.28}, \eqref{2.29}, and initial and boundary conditions Eqs.\eqref{2.30}-\eqref{2.36}. All equations, initial and boundary conditions are summarized in Table \ref{t2.1}. The reaction terms of these equations depend on the specific biological case considered and describe the complex biological interplay taking place between sessile biomass $X_i(r,t)$, planktonic biomass $\psi_i(r,t)$ and soluble substrates $S_j(r,t)$ within the biofilm and planktonic biomass $\psi^*_i(t)$ and soluble substrates $S^*_j(t)$ within the bulk liquid.

\section{Modelling \textit{de novo} anaerobic granulation} 
\label{n3}
    
The mathematical model described in the previous section can be applied to any granular biofilm system by defining appropriate variables, parameters, initial and boundary conditions and reaction terms based on the biological processes involved.

In this work, the model is applied to study the process of \textit{de novo} granulation and the ecology of granules in an anaerobic reactor. Anaerobic digestion (AD) is a biological process extensively used to manage liquid and solid waste and produce fuels. It is a low-cost and renewable technology based on a complex multi-step process where different microbial species convert organic substances in methane-rich biogas. During the last decades, the AD process has been implemented in granular biofilm systems, where the microbial community forms dense biofilm granules which offer several operational advantages compared to conventional suspended systems \cite{baeten2019modelling,liu2002essential}.

Many studies report the fundamental role played by methanogenic species which facilitate the formation of the granule nucleus due to their filamentous structure and their abilities to produce EPS and to use quorum sensing strategies \cite{trego2019granular, li2015significant, zhang2012acyl, li2014characterization}. To model this aspect, different attachment velocities are used depending on the microbial species.

The following variables, expressed in terms of concentrations, are included in the model:

\begin{itemize}
 \item 5 sessile microbial species: sugar fermenters $X_{Su}$, butyrate consumers $X_{Bu}$, propionate consumers $X_{Pro}$, acetoclastic methanogens $X_{Ac}$, inert material $X_{I}$.
 \item 4 planktonic species within the biofilm: sugar fermenters $\psi_{Su}$, butyrate consumers $\psi_{Bu}$, propionate consumers $\psi_{Pro}$, acetoclastic methanogens $\psi_{Ac}$.
 \item 5 soluble substrates within the biofilm: sugar $S_{Su}$, butyrate $S_{Bu}$, propionate $S_{Pro}$, acetate $S_{Ac}$ and methane $S_{CH_4}$.
 \item 4 planktonic species within the bulk liquid: sugar fermenters $\psi^*_{Su}$, butyrate consumers $\psi^*_{Bu}$, propionate consumers $\psi^*_{Pro}$, acetoclastic methanogens $\psi^*_{Ac}$.
 \item 5 soluble substrates within the bulk liquid: sugar $S^*_{Su}$, butyrate $S^*_{Bu}$, propionate $S^*_{Pro}$, acetate $S^*_{Ac}$ and methane $S^*_{CH_4}$.
 \end{itemize} 

Inert material is not considered in the bulk liquid as it is supposed to play no role in the life cycle of the granular biofilm (inactive biomass is supposed to have neither metabolic activity nor granulation and invasion properties).

The model considers an influent flow composed exclusively of dissolved substrates. Therefore, disintegration and hydrolysis processes, which lead to the conversion of organic matter into soluble compounds, are neglected. The main intracellular processes are taken into account both in the biofilm and in the bulk liquid: acidogenesis, acetogenesis and methanogenesis. The kinetic expressions of the biological processes involved in the model are taken from \cite{batstone2002iwa}. In particular, each growth process leads to the formation of new biomass, consumption and/or production of one or more soluble substrates. Each decay process implies the death of active biomass which turns into inert material. Within the biofilm, sessile sugar fermenters $X_{Su}$ grow by converting sugar $S_{Su}$ into butyrate $S_{Bu}$, propionate $S_{Pro}$ and acetate $S_{Ac}$ (acidogenesis process). Butyrate $S_{Bu}$ and propionate $S_{Pro}$ are consumed by sessile butyrate consumers $X_{Bu}$ and sessile propionate consumers $X_{Pro}$, respectively, and acetate $S_{Ac}$ is produced (acetogenesis process). Lastly, acetate is converted into methane $S_{CH_4}$ by sessile acetoclastic methanogens $X_{Ac}$ (methanogenesis process). The same biological processes are supposed to occur in the bulk liquid where the planktonic biomasses $\psi^*_i$ consume or produce the soluble substrates $S^*_j$. Furthermore, the decay of any sessile biomass is considered to produce inert material $X_I$ which represents inactive biomass and accumulates in the biofilm. The decay processes are also considered for planktonic species in the bulk liquid.

The planktonic active species present in the bulk liquid are also modelled in the granule domain as planktonic cells $\psi_i$ which populate the voids of the solid matrix and contribute to the growth of the corresponding sessile species as a result of the invasion phenomena.

All the reaction terms of the model equations are listed below. The specific growth rates within the biofilm due to sessile biomass $r_{M,i}$ in Eq.\eqref{2.10} and Eq.\eqref{2.12} are modelled as Monod-type kinetics:

\begin{equation}      \label{3.1}                                        
r_{M,i} = f_{i} (\mu_{\max,i} \frac{S_{i}}{K_{i}+S_{i}}-k_{d,i}), \ i \in I_B,
\end{equation}

where $I_B = \{Su, Bu, Pro, Ac\}$ is the index set, $\mu_{\max,i}$ is the maximum net growth rate for biomass $i$, $K_i$ is the affinity constant of the consumed substrate for biomass $i$ and $k_{d,i}$ is the decay constant for biomass $i$.

The inert formation rate is give by the sum of the decay rates of each active species, modelled as first order kinetic:

 \begin{equation}                                          \label{3.2}
r_{M,I} = \sum_{i \in I_B} f_{i} \ k_{d,i}.
\end{equation}
\\

The specific growth rates within the biofilm due to planktonic cells $r_i$ in Eq.\eqref{2.10} and Eq.\eqref{2.12} are defined as:  

\begin{equation}                                          \label{3.3}
r_{i} = k_{col,i} \ \frac{\psi_{i}}{\rho} \frac{S_{i}}{K_{i}+S_{i}}, \ i \in I_B,
\end{equation}

where $k_{col,i}$ is the maximum colonization rate of motile species $i$ and ${\rho}$ is the granule density.

The conversion rates for planktonic cells due to the invasion process $r_{\psi,i}$ in Eq.\eqref{2.15} are expressed by:

\begin{equation}                                          \label{3.4}
r_{\psi,i} = -\frac{1}{Y_{\psi,i}}r_{i} \ \rho, \ i \in I_B,
\end{equation}

where $Y_{\psi,i}$ denotes the yield of non-motile species $i$ on corresponding motile species. 

The conversion rates for soluble substrates within the biofilm $r_{S,j}$ in Eq.\eqref{2.16}, with $j \in \{Su,Bu,Pro,Ac,CH_4\}$, are listed below:

\begin{equation}                                          \label{3.5}
r_{S,Su} = - \frac{\mu_{\max,Su}}{Y_{Su}} \frac{S_{Su}}{K_{Su} + S_{Su}}f_{Su} \ \rho,
\end{equation}	

\[
r_{S,{Bu}} = - \frac{\mu_{\max,{Bu}}}{Y_{Bu}} \frac{S_{Bu}}{K_{{Bu}}  +S_{Bu}}f_{Bu} \ \rho \ + \ g_{Su,Bu} \frac{(1-Y_{Su})}{Y_{Su}} \mu_{\max,Su} \times 
\]  
\begin{equation}                                          \label{3.6}
\times \frac{S_{Su}}{K_{Su}+S_{Su}}f_{Su} \ \rho  ,
\end{equation}	
  
\[
r_{S,{Pro}} = - \frac{\mu_{\max,{Pro}}}{Y_{Pro}} \frac{S_{Pro}}{K_{{Pro}}+S_{Pro}}f_{Pro} \ \rho \ + \ g_{Su,Pro} \frac{(1-Y_{Su})}{Y_{Su}} \mu_{\max,Su} \times 
\]
\begin{equation}                                          \label{3.7}
\times \frac{S_{Su}}{K_{Su}+S_{Su}}f_{Su} \ \rho  ,
\end{equation}	
  
\[
r_{S,{Ac}} = -\frac{\mu_{\max,{Ac}}}{Y_{Ac}} \frac{S_{Ac}}{K_{{Ac}}+S_{Ac}}f_{Ac} \ \rho \ + \ g_{Su,{Ac}} \frac{(1-Y_{Su})}{Y_{Su}}\mu_{\max,Su} \frac{S_{Su}}{K_{Su}+S_{Su}} \times
\]
\[
\times f_{Su} \ \rho  \ + g_{Bu,Ac} \frac{(1-Y_{Bu})}{Y_{Bu}} \mu_{\max,{Bu}} \frac{S_{Bu}}{K_{{Bu}}+S_{Bu}}f_{Bu} \ \rho \ + \ g_{Pro,Ac} \frac{(1-Y_{Pro})}{Y_{Pro}} \times
\]	
 \begin{equation}                                         \label{3.8}
 \times \mu_{\max,{Pro}} \frac{S_{Pro}}{K_{{Pro}}+S_{Pro}}f_{Su} \ \rho  ,
 \end{equation}

\begin{equation}                                          \label{3.9}
r_{S,{CH_4}} = \frac{(1-Y_{Ac})}{Y_{Ac}} \mu_{\max,{Ac}} \frac{S_{Ac}}{K_{{Ac}}+S_{Ac}}f_{Ac} \ \rho, 
\end{equation}

where $Y_{Su}$, $Y_{Bu}$, $Y_{Pro}$, $Y_{Ac}$, denote the yields of sugar fermenters, butyrate consumers, propionate consumers and acetoclastic methanogens on the corrisponding substrate consumed, $g_{Su,Bu}$, $g_{Su,Pro}$, $g_{Su,Ac}$ are the stoichiometric fractions of butyrate, propionate and acetate produced from sugar, $g_{Bu,Ac}$ and $g_{Pro,Ac}$ are the stoichiometric fractions of acetate produced from butyrate and propionate.

 \begin{table}[ht]
\begin{tiny}
\begin{spacing}{1.2}
 \begin{center}
 \begin{tabular}{llccc}
 \hline
{\textbf{Parameter}} & {\textbf{Definition}} & {\textbf{Unit}} & {\textbf{Value}} & {\textbf{Ref}}
 \\
 \hline
 $\mu_{max,Su}$ & Maximum specific growth rate for sugar fermenters &  $d^{-1}$ & $3$ & (a) \\
 $\mu_{max,Bu}$ & Maximum specific growth rate for butyrate consumers &  $d^{-1}$  & $1.2$  & (a) \\
  $\mu_{max,Pro}$ & Maximum specific growth rate for propionate consumers  &  $d^{-1}$ & $0.52$  & (a)\\
  $\mu_{max,Ac}$ & Maximum specific growth rate for acetoclastic methanogens &  $d^{-1}$  & $0.4$ & (a) \\
  
 $k_{d,Su}$ & Decay-inactivation rate for sugar fermenters  &  $d^{-1}$ & $0.02$  & (a) \\
 $k_{d,Bu}$  & Decay-inactivation rate for butyrate consumers &  $d^{-1}$ & $0.02$  & (a) \\
 $k_{d,Pro}$ & Decay-inactivation rate for propionate consumers        &  $d^{-1}$  & $0.02$  & (a) \\
 $k_{d,Ac}$ & Decay-inactivation rate for acetoclastic methanogens        &  $d^{-1}$ & $0.02$  & (a) \\
 
 $K_{Su}$    & Sugar half saturation constant sugar fermenters &  $gCOD \ m^{-3}$ & $500$ & (a) \\
 $K_{Bu}$ & Butyrate half saturation constant butyrate consumers  &  $gCOD\ m^{-3}$ & $300$ & (a) \\
 $K_{Pro}$ & Propionate half saturation constant propionate consumers &  $gCOD \ m^{-3}$ & $300$ &(a) \\
 $K_{Ac}$ & Acetate half saturation constant acetoclastic methanogens & $gCOD \ m^{-3}$ & $150$ & (a) \\
 
 $Y_{Su}$ & Yield of sugar fermenters on sugar  & $--$ & $0.10$ & (a) \\ 
 $Y_{Bu}$ & Yield of butyrate consumers on butyrate  & $--$ & $0.06$ & (a)  \\ 
 $Y_{Pro}$ & Yield of propionate consumers on propionate  & $--$ & $0.04$ & (a)  \\ 
 $Y_{Ac}$ & Yield of acetoclastic methanogens on acetate & $--$    & $0.05$ & (a) \\ 
 
 $g_{Su,Bu}$ & Fraction of butyrate from sugar  &  $--$    & $0.13$ & (a) \\ 
 $g_{Su,Pro}$ & Fraction of propionate from sugar & $--$ & $0.27$ & (a) \\ 
 $g_{Su,Ac}$ & Fraction of acetate from sugar & $--$ & $0.41$ &   (a) \\ 
 $g_{Bu,Ac}$ & Fraction of acetate from butyrate & $--$  & $0.80$    & (a) \\  
 $g_{Pro,Ac}$ & Fraction of acetate from propionate & $--$ & $0.57$ & (a) \\  
 
 $D_{S,Su}$ & Diffusion coefficient of sugar in biofilm & 
 $m^2 \ d^{-1}$ & $4.63\cdot10^{-5}$ & (b) \\
 $D_{S,Bu}$ & Diffusion coefficient of butyrate in biofilm & $m^2 \ d^{-1}$ & $6.01\cdot10^{-5}$ & (b) \\
 $D_{S,Pro}$ & Diffusion coefficient of propionate in biofilm & $m^2 \ d^{-1}$ & $7.33\cdot10^{-5}$ &  (b) \\
 $D_{S,Ac}$ & Diffusion coefficient of acetate in biofilm & $m^2 \ d^{-1}$ & $8.36\cdot10^{-5}$ & (b) \\
 $D_{S,CH_4}$ & Diffusion coefficient of methane in biofilm & $m^2 \ d^{-1}$ & $10.3\cdot10^{-5}$ & (b) \\
 
  $k_{col,i}$ & Maximum colonization rate of $i^{th}$ planktonic species & $d^{-1}$ & $0.001$  & (c)  \\
  
  $Y_{\psi,i}$ & Yield of non-motile microorganisms on motile species   & $--$ & $0.001$ & (c) \\

  $D_{\psi,i}$ & Diffusion coefficient of $i^{th}$ planktonic species in biofilm  & $m^2 \ d^{-1}$ & $10^{-5}$ &  (c)  \\
    
  $v_{a,Su}$    & Attachment velocity of planktonic species $\psi_{Su}$ & $m \ d^{-1}$ & $3\cdot10^{-3}$    &    (c) \\
   $v_{a,Bu}$    & Attachment velocity of planktonic species $\psi_{Bu}$ & $m \ d^{-1}$ & $3\cdot10^{-3}$    &    (c)    \\
   $v_{a,Pro}$    & Attachment velocity of planktonic species $\psi_{Pro}$ & $m \ d^{-1}$ & $3\cdot10^{-3}$    &    (c)    \\
   $v_{a,Ac}$    & Attachment velocity of planktonic species $\psi_{Ac}$ & $m \ d^{-1}$  & $150\cdot10^{-3}$    &    (c)    \\
 $\rho$    & Biofilm density  & $gCOD \ m^{-3}$ & $120000$  &    (c)  \\
 
 $\lambda$    & Detachment coefficient   &  $m^{-1} \ d^{-1}$ & 10  &    (c)   \\
  $V$  & Reactor volume  &  $m^{3}$   & 400  &    (c)   \\
  $Q$  &  Volumetric flow rate &  $m^{3} \ d^{-1}$  & 600  &    (c)   \\
  $N_G$  &  Number of granules in the reactor &  $--$   & $2.4\cdot10^{10}$ & (c)   \\
 \hline
 \multicolumn{5}{l}{(a) \cite{batstone2002iwa}; (b) \cite{stewart2003diffusion}; (c) Assumed}\\
 \end{tabular}
 \caption{Kinetic, stoichiometric and operating parameters used for numerical simulations} \label{t3.1}
 \end{center}
 \end{spacing}
 \end{tiny}
 \end{table}

Moreover, the conversion rates of planktonic biomasses $r^*_{\psi,i}$ within the bulk liquid in Eq.\eqref{2.28} are defined as:

\begin{equation}                                          \label{3.10}
r^*_{\psi,i} = \psi_{i}^* (\mu_{\max,i} \frac{S_{i}^*}{K_{i}+S_{i}^*}-k_{d,i}), \ i \in I_B,
\end{equation}

while, the conversion rates of soluble substrates $r^*_{S,j}$ within the bulk liquid in Eq.\eqref{2.29}, with $j \in \{Su,Bu,Pro,Ac,CH_4 \}$, are listed below:

\begin{equation}                                          \label{3.11}
r^*_{S,Su} = -\psi_{Su}^* \frac{\mu_{\max,Su}}{Y_{Su}} \frac{S_{Su}^*}{K_{Su}+S_{Su}^*},
\end{equation}	
  
\[
r^*_{S,{Bu}} = -\psi_{Bu}^* \frac{\mu_{\max,{Bu}}}{Y_{Bu}} \frac{S_{Bu}^*}{K_{{Bu}}+S_{Bu}^*} + g_{Su,Bu} \frac{(1-Y_{Su})}{Y_{Su}} \psi_{Su}^* \mu_{\max,Su} \times
\]
\begin{equation}                                          \label{3.12}
 \times \frac{S_{Su}^*}{K_{Su}+S_{Su}^*} ,
\end{equation}	
  
\[
r^*_{S,{Pro}} = -\psi_{Pro}^*  \frac{\mu_{\max,{Pro}}}{Y_{Pro}} \frac{S_{Pro}^*}{K_{{Pro}}+S_{Pro}^*} \ + \ g_{Su,Pro} \frac{(1-Y_{Su})}{Y_{Su}} \psi_{Su}^* \mu_{\max,Su} \times
\]
\begin{equation}                                          \label{3.13}
 \times \frac{S_{Su}^*}{K_{Su}+S_{Su}^*},
\end{equation}	
    
\[
r^*_{S,{Ac}} = -\psi_{Ac}^* \frac{\mu_{\max,{Ac}}}{Y_{Ac}} \frac{S_{Ac}^*}{K_{{Ac}}+S_{Ac}^*} + g_{Su,Ac} \frac{(1-Y_{Su})}{Y_{Su}} \psi_{Su}^* \mu_{\max,Su} \frac{S_{Su}^*}{K_{Su}+S_{Su}^*}  \ + \]
\[
 g_{Bu,Ac} \frac{(1-Y_{Bu})}{Y_{Bu}} \psi_{Bu}^* \mu_{\max,{Bu}} \frac{S_{Bu}^*}{K_{{Bu}}+S_{Bu}^*} \ + \ g_{Pro,Ac} \frac{(1-Y_{Pro})}{Y_{Pro}} \psi_{Pro}^* \mu_{\max,{Pro}} \times
 \]
 \begin{equation}                                          \label{3.14}
\times \frac{S_{Pro}^*}{K_{{Pro}}+S_{Pro}^*},
\end{equation}

\begin{equation}                                          \label{3.15}
r^*_{S,{CH_4}} = \frac{(1-Y_{Ac})}{Y_{Ac}} \psi_{Ac}^* \mu_{\max,{Ac}} \frac{S_{Ac}^*}{K_{{Ac}}+S_{Ac}^*}. 
\end{equation}

The values used for all stoichiometric and kinetic parameters are reported in Table \ref{t3.1}.

\section{Numerical simulations and results} 
\label{n4}

The model has been integrated numerically by developing an original code in MatLab platform, essentially based on the method of characteristics and the method of lines.

Numerical simulations are performed to describe the formation and evolution of anaerobic granular biofilms, to study the ecological succession occurring in the granule and explore the effects of the main factors on the process. In particular, five numerical studies are carried out: the first numerical study (NS1) describes the \textit{de novo} granulation process in a reactor fed with an influent wastewater rich in sugar; the second study (NS2) investigates the effect of the influent composition on the granule evolution and ecology; the third study (NS3) explores the role of the attachment phenomenon on the granule evolution; the fourth study (NS4) investigates the effects of the biomass density on the transport of soluble substrates  and, consequently, on the growth and stratification of biomass within the granule; lastly, the fifth study (NS5) simulates the effects of different detachment regimes on granule dimension and dynamics. The values used for the parameters under study are presented in Table \ref{t4.1} for all numerical studies.

 \begin{table}[ht]
 \begin{tiny}
 \begin{spacing}{1.2}
 \begin{center}
 \begin{tabular}{lcccccc}
 \hline
\multirow{2}*{\textbf{Parameter}} & {\textbf{NS1}} & {\textbf{NS2}} & {\textbf{NS3}} & {\textbf{NS4}} & {\textbf{NS5}}\\
\cmidrule(l){2-6}
& {\text{RUN1}} & {\text{RUN2 - RUN4}} & {\text{RUN5 - RUN13}} & {\text{RUN14 - RUN17}} & {\text{RUN18 - RUN25}}
\\ 
\hline
 $S^{in}_{Su}$ $[gCOD \ m^{-3}]$ & $3500$ & $varied^1$ & $3500$ & $3500$ & $3500$\\
 $S^{in}_{Bu}$ $[gCOD \ m^{-3}]$ & $0$ & $varied^1$ & $0$ & $0$ & $0$ \\ 
 $S^{in}_{Pro}$ $[gCOD \ m^{-3}]$ & $0$ & $varied^1$ & $0$ & $0$ & $0$ \\
 $S^{in}_{Ac}$ $[gCOD \ m^{-3}]$ & $0$ & $varied^1$ & $0$ & $0$ & $0$ \\
 $S^{in}_{CH_4}$ $[gCOD \ m^{-3}]$ & $0$ &    $varied^1$ &    $0$ & $0$ & $0$  \\
 $\psi^{*}_{Su,0}$ $[gCOD \ m^{-3}]$ & $300$ & $varied^1$ & $300$ & $300$ & $300$ \\
 $\psi^{*}_{Bu,0}$ $[gCOD \ m^{-3}]$ & $50$  & $varied^1$ & $50$ & $50$ & $50$  \\
 $\psi^{*}_{Pro,0}$ $[gCOD \ m^{-3}]$ & $50$  & $varied^1$  & $50$ & $50$ & $50$  \\
 $\psi^{*}_{Ac,0}$ $[gCOD \ m^{-3}]$ & $100$ & $varied^1$ & $100$ & $100$ & $100$ \\
 $v_{a,Su}$ $[m \ d^{-1}]$  & $3\cdot10^{-3}$  & $3\cdot10^{-3}$ & $varied^1$ &  $3\cdot10^{-3}$ & $3\cdot10^{-3}$ \\
 $v_{a,Bu}$ $[m \ d^{-1}]$  & $3\cdot10^{-3}$  & $3\cdot10^{-3}$ & $varied^1$ & $3\cdot10^{-3}$ & $3\cdot10^{-3}$ \\
 $v_{a,Pro}$ $[m \ d^{-1}]$  & $3\cdot10^{-3}$  & $3\cdot10^{-3}$ & $varied^1$ & $3\cdot10^{-3}$ & $3\cdot10^{-3}$ \\
 $v_{a,Ac}$ $[m \ d^{-1}]$  & $150\cdot10^{-3}$  & $150\cdot10^{-3}$ & $varied^1$ & $150\cdot10^{-3}$ & $150\cdot10^{-3}$ \\
 $\rho$ $[gCOD \ m^{-3}]$ & $120000$ & $120000$ & $120000$ & $varied^1$ & $120000$ \\
 $\lambda$ $[m^{-1} \ d^{-1}]$  & $10$ & $10$ & $10$ & $10$ & $varied^1$ \\
 $T$ $[d] $ & 300 & 300 & 300 & 300 & 300 \\
 \hline
 \multicolumn{6}{l}{$^1$The values used are reported in the text} \\
 \end{tabular}
 \caption{Initial and boundary conditions and operating parameters adopted in numerical studies} \label{t4.1}
 \end{center}
 \end{spacing}
 \end{tiny}
 \end{table}

The initial concentration of the soluble substrates in the bulk liquid $S^*_{j,0}$ is assumed the same as the influent wastewater. No microbial biomass is present in the influent flow ($\psi^{in}_i=0$), while non-null initial concentrations of planktonic biomasses in the bulk liquid $\psi^*_{i,0}$ are set to simulate the reactor inoculated with an anaerobic sludge. In particular, it is considered that the granular reactor is inoculated with the sludge coming from a conventional AD reactor and fed with the same influent wastewater. Therefore, the initial concentrations of planktonic species in the bulk liquid (representative of the inoculum) are derived from numerical results of an ADM1-based model \cite{batstone2002iwa}. 

In granular reactors, intense hydrodynamic conditions shall improve the aggregation of planktonic cells and the formation of granules. Consequently, for all the simulations reported in this work, the reactor volume $V$ and the influent flow rate $Q$ are assumed constant and equal to $400 \ m^3$ and $600 \ m^3 d^{-1}$, respectively, leading to high hydrodynamic velocities and a very low hydraulic retention time (HRT = $0.667 \ d$). Such value is within the range of HRT values typical of granular biofilm systems \cite{lim2014applicability}. Moreover, the organic loading rate (OLR), defined as the amount of daily organic matter treated per unit reactor volume, is set equal to $5.25 \ kg \  m^{-3} \ d^{-1}$. The number of granules $N_G$ is calculated to guarantee a $25\%$ filling ratio of the reactor volume at steady-state condition.

Diffusivity of soluble substrates in biofilm is assumed to be $80\%$ of diffusivity in water \cite{wanner1986multispecies}. The diffusion coefficients in water for all soluble substrates are taken from \cite{stewart2003diffusion}, see Table \ref{t3.1}. 

 The simulation time $T$ is fixed to $300 \ d$ for all simulations. This time interval guarantees to achieve the steady-state configuration for all model variables: concentration of soluble substrates $S^*_j(t)$ and planktonic biomasses $\psi^*_i(t)$ in the bulk liquid; granule dimension $R(t)$; sessile biomass fractions $f_i(r,t)$, concentration of soluble substrates $S_j(r,t)$ and planktonic species $\psi_i(r,t)$ within the biofilm.

\subsection{NS1 - Anaerobic granulation process} 
\label{n4.1}
 The first numerical study (NS1) describes the \textit{de novo} granulation process occurring in a granular reactor fed with sugar: $S^{in}_{Su} = 3500 \ g \  m^{-3}$, $S^{in}_{Bu} = S^{in}_{Pro} = S^{in}_{Ac} = S^{in}_{CH_4} = 0$ (RUN1). The initial concentration of the planktonic biomasses (reactor inoculum) is derived from an ADM1-based model following the procedure introduced above: $\psi^*_{Su,0} = 300 \ g \  m^{-3}$, $\psi^*_{Bu,0} = \psi^*_{Pro,0} = 50 \ g \  m^{-3}$, $\psi^*_{Ac,0} = 100 \ g \  m^{-3}$. 
 
\begin{figure*}
 \fbox{\includegraphics[width=1\textwidth, keepaspectratio]{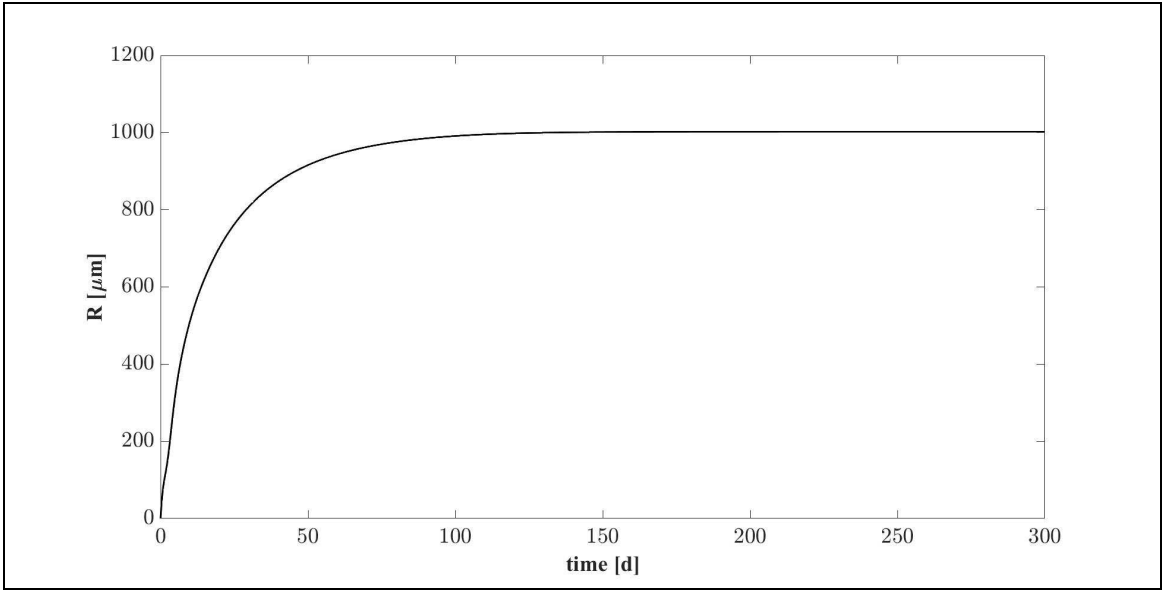}}   
 \caption{NS1 - Biofilm radius evolution over time. Influent wastewater composition: $S^{in}_{Su} = 3500 \ g \ m^{-3}$ (Sugar), $S^{in}_{Bu} = 0$ (Butyrate), $S^{in}_{Pro} = 0$ (Propionate), $S^{in}_{Ac} = 0$ (Acetate), $S^{in}_{CH_4} = 0$ (Methane)}
          \label{f4.1.1} 
 \end{figure*}

\begin{figure*}
 \fbox{\includegraphics[width=1\textwidth, keepaspectratio]{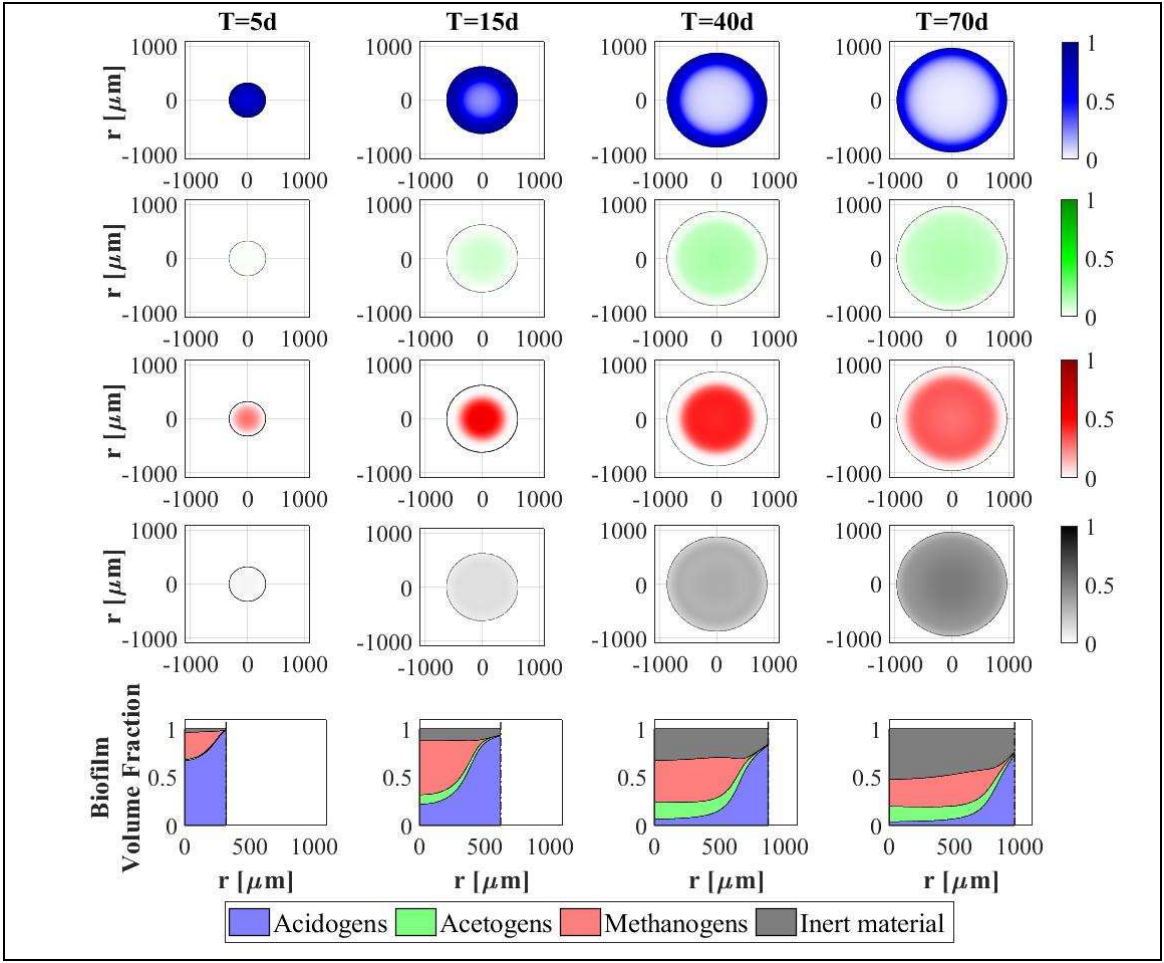}}   
 \caption{NS1 - Microbial species distribution in the diametrical section and across the radius of the granule, at $T = 5 \ d$, $T = 15 \ d$, $T = 40 \ d$ and $T = 70 \ d$. Influent wastewater composition: $S^{in}_{Su} = 3500 \ g \ m^{-3}$ (Sugar), $S^{in}_{Bu} = 0$ (Butyrate), $S^{in}_{Pro} = 0$ (Propionate), $S^{in}_{Ac} = 0$ (Acetate), $S^{in}_{CH_4} = 0$ (Methane)}
          \label{f4.1.2} 
 \end{figure*}

\begin{figure*}
 \fbox{\includegraphics[width=1\textwidth, keepaspectratio]{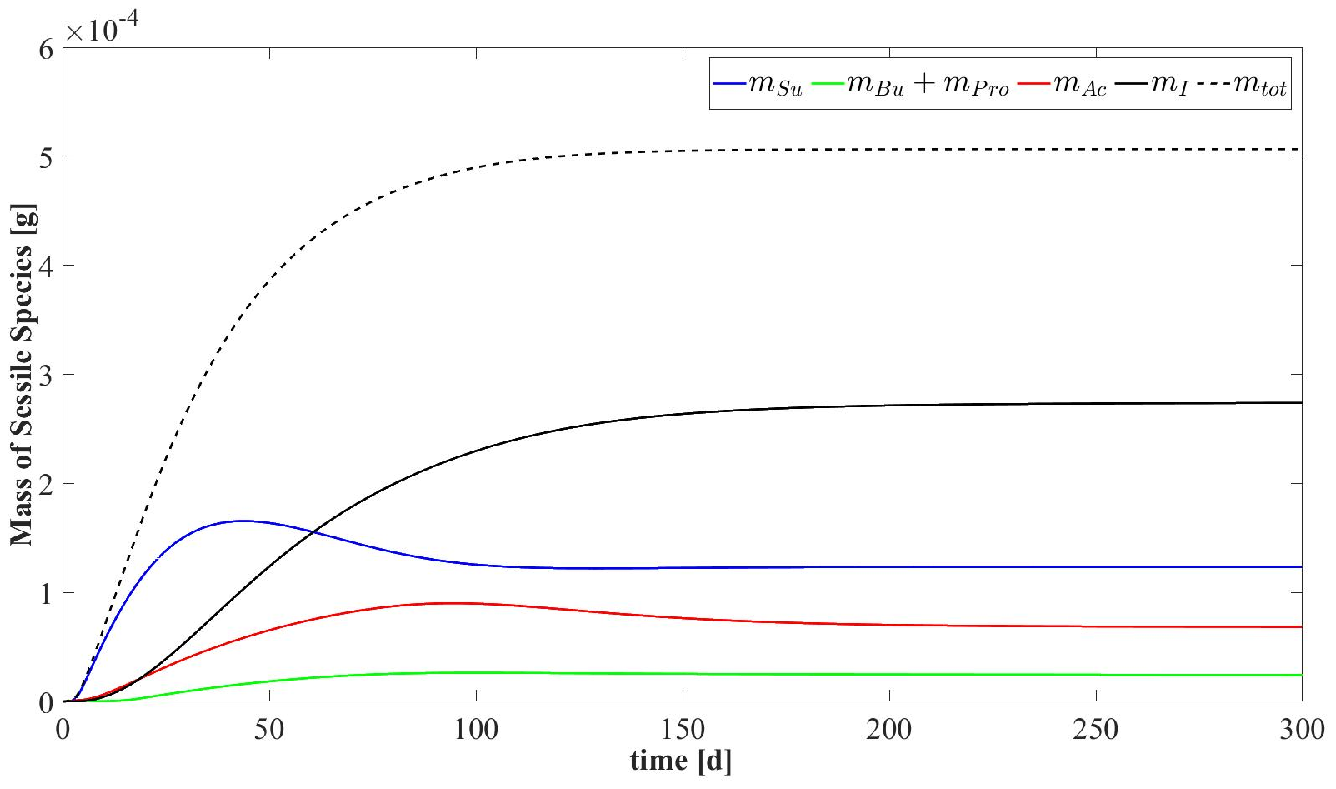}}   
 \caption{NS1 - Evolution of mass of sessile species within the granule. $m_{Su}$: mass of sugar fermenters, $m_{Bu}$: mass of butyrate consumers, $m_{Pro}$: mass of propionate consumers, $m_{Ac}$: mass of acetoclastic methanogens, $m_{tot}$: total sessile mass. Influent wastewater composition: $S^{in}_{Su} = 3500 \ g \ m^{-3}$ (Sugar), $S^{in}_{Bu} = 0$ (Butyrate), $S^{in}_{Pro} = 0$ (Propionate), $S^{in}_{Ac} = 0$ (Acetate), $S^{in}_{CH_4} = 0$ (Methane)}
          \label{f4.1.3} 
 \end{figure*}

\begin{figure*}
 \fbox{\includegraphics[width=1\textwidth, keepaspectratio]{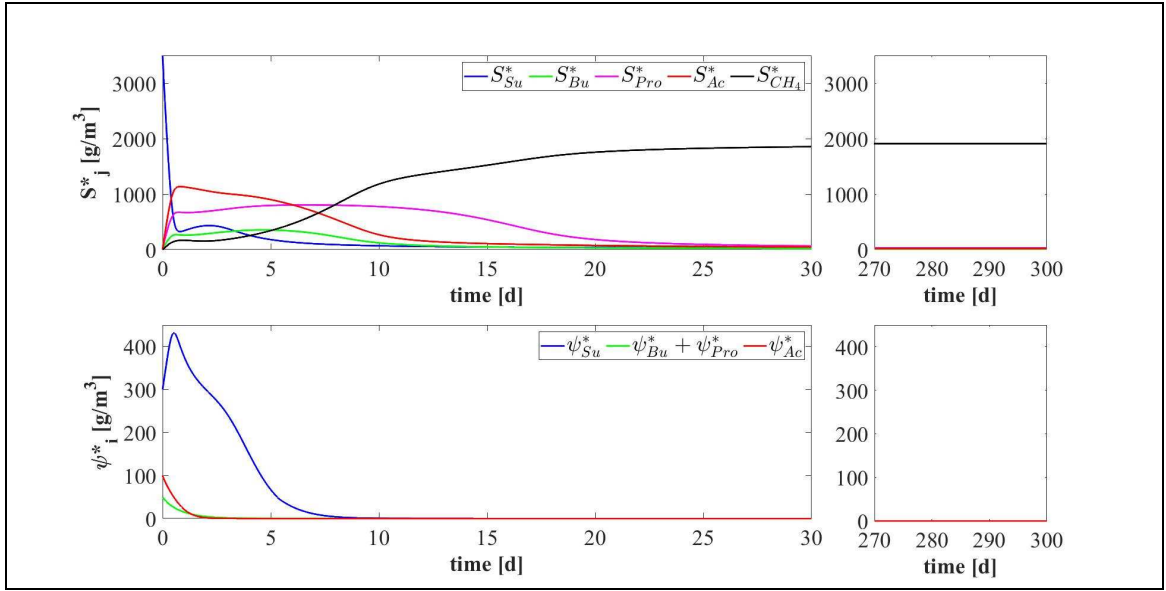}}   
 \caption{NS1 - Evolution of soluble substrates (top) and planktonic biomass (bottom) concentrations within the bulk liquid. $S^*_{Su}$: Sugar, $S^*_{Bu}$: Butyrate, $S^*_{Pro}$: Propionate, $S^*_{Ac}$: Acetate, $S^*_{CH_4}$: Methane, $\psi^*_{Su}$: Sugar fermenters, $\psi^*_{Bu}$: Butyrate consumers, $\psi^*_{Pro}$: Propionate consumers, $\psi^*_{Ac}$: Acetoclastic methanogens. Influent wastewater composition: $S^{in}_{Su} = 3500 \ g \ m^{-3}$ (Sugar), $S^{in}_{Bu} = 0$ (Butyrate), $S^{in}_{Pro} = 0$ (Propionate), $S^{in}_{Ac} = 0$ (Acetate), $S^{in}_{CH_4} = 0$ (Methane)}
          \label{f4.1.4} 
 \end{figure*}

Numerical results are summarized in Figs. \ref{f4.1.1}-\ref{f4.1.4}. In Fig. \ref{f4.1.1} the evolution of the granule radius $R(t)$ over time is reported. A vanishing initial value is assigned to $R(t)$ at $t=0$ ($R(0) = 0$). During the first days, the granulation process has its maximum intensity and the granule dimension increases. The variation of $R(t)$ is almost exhausted during the first $70$-$100$ days, after which it reaches a steady-state value of about $1 \ mm$.

In Fig. \ref{f4.1.2} the distribution of sessile species within the granule is shown at different times. After $5 \ d$ the granule has a radius of about $0.3 \  mm$ and is constituted mostly by acidogens (blue) which are favoured by the high concentration of sugar initially present in the bulk liquid. However, the granule core is also composed of methanogens (red) which have high propensity to attach due to their filamentous structures and aggregation properties. At $T=15 \ d$, the consumption of sugar and the production of volatile fatty acids (VFAs) in large amount by acidogenesis affects the biomass distribution: the methanogenic core grows while the acidogens occupy the outer layer of the granule, and the acetogens (green) and inert (black) fractions starts to be visible. For later times ($40$-$70 \ d$), the radius almost reaches the steady-state value, a significant amount of inert material accumulates especially in the innermost part of the domain, homogeneous fractions of methanogens and acetogens are found throughout the granule except the outermost part, where a thin layer of acidogens is established.

 Fig. \ref{f4.1.3} presents the trend of each microbial species within the granule over time. This result confirms the microbial succession described above. The biofilm is initially constituted predominantly by acidogens $m_{Su}$ (blue) and methanogens $m_{Ac}$ (red). Their mass within the biofilm achieves a maximum and then decreases to a steady-state value when the substrates required by their metabolism (sugar and acetate, respectively) are limited and the decay process prevails. The growth process of acetogens $m_{Bu}+m_{Pro}$ and the accumulation of inert $m_{I}$ are slower and take place in a longer time. However, all microbial species exhibit steady-state values $150$-$170$ days after the granule genesis. Furthermore, the total microbial mass $m_{tot}$ (dashed black line) follows the trend of the radius reported in Fig. \ref{f4.1.1}. Indeed, assuming a constant density $\rho$, the variation of mass within the granule is related to the variation of volume.
 
Lastly, the trends of soluble substrates and planktonic biomass within the bulk liquid are shown in Fig. \ref{f4.1.4}. In the initial phase, the biofilm granules are small, and the consumption and production of soluble substrates are governed by planktonic biomass. In particular, planktonic sugar fermenters $\psi^*_{Su}$ (blue in Fig. \ref{f4.1.4}-bottom) degrade sugar $S^*_{Su}$ (blue in Fig. \ref{f4.1.4}-top) and produce VFAs: butyrate $S^*_{Bu}$ (green in Fig. \ref{f4.1.4}-top), propionate $S^*_{Pro}$ (magenta in Fig. \ref{f4.1.4}-top) and acetate $S^*_{Ac}$ (red in Fig. \ref{f4.1.4}-top). Meanwhile, the concentration of all planktonic biomasses within the bulk liquid is reduced due to two distinct phenomena: part is converted in sessile biomass during the granulation process and part is rapidly washed out due to the hydrodynamic conditions (low HRT). For these reasons, no microbial species in planktonic form is present within the reactor after $5$-$7$ days. After the washout of the planktonic biomass, the substrates trend is influenced exclusively by the sessile metabolic activity: the residual sugar $S^*_{Su}$ and VFAs ($S^*_{Bu}$, $S^*_{Pro}$ and $S^*_{Ac}$) are consumed with different velocities according to the consumption rate of the corresponding sessile microbial species and significant amount of methane $S^*_{CH_4}$ (black in Fig. \ref{f4.1.4}-top) is produced. After $30$ days substrates concentrations within the bulk liquid reach a steady-state value. High concentrations of methane (end product of the AD process) and negligible concentrations of sugar and VFAs are found inside the reactor and in the effluent.

\subsection{NS2 - Effects of influent wastewater composition} 
\label{n4.2}

\begin{figure*}
 \fbox{\includegraphics[width=1\textwidth, keepaspectratio]{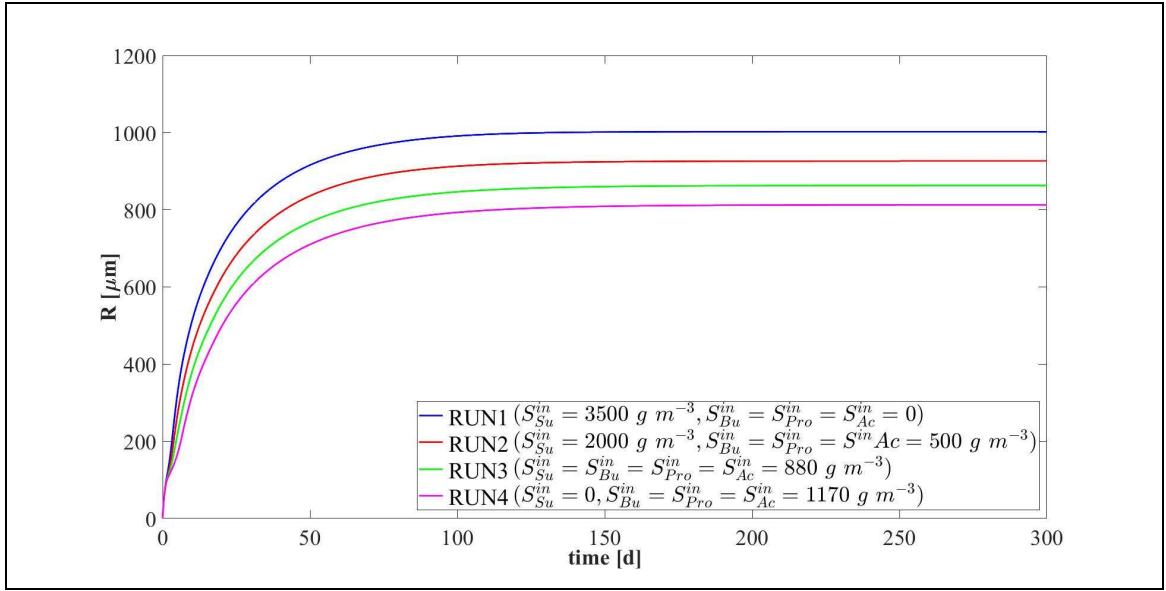}}   
 \caption{NS2 - Biofilm radius evolution over time for different influent wastewater compositions. $S^{in}_{Su}$: Sugar, $S^{in}_{Bu}$: Butyrate, $S^{in}_{Pro}$: Propionate, $S^{in}_{Ac}$: Acetate, $S^{in}_{CH_4}$: Methane}
          \label{f4.2.1} 
 \end{figure*}

\begin{figure*}
 \fbox{\includegraphics[width=1\textwidth, keepaspectratio]{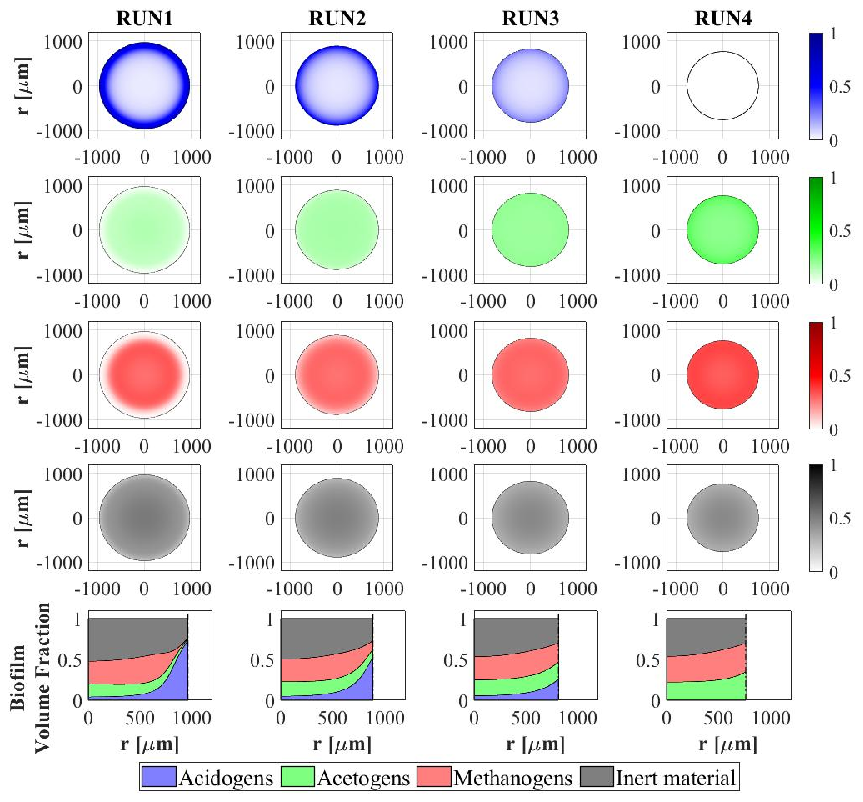}}   
 \caption{NS2 - Microbial species distribution in the diametrical section and across the radius of the granule at $T = 70 \ d$, for different influent wastewater compositions. RUN1: $S^{in}_{Su} = 3500 \ g \ m^{-3}$, $S^{in}_{Bu} = S^{in}_{Pro} = S^{in}_{Ac} = S^{in}_{CH_4} = 0$; RUN2: $S^{in}_{Su} = 2000 \ g \ m^{-3}$, $S^{in}_{Bu} = S^{in}_{Pro} = S^{in}_{Ac} = 500 \ g \ m^{-3}$, $S^{in}_{CH_4} = 0$; RUN3: $S^{in}_{Su} = S^{in}_{Bu} = S^{in}_{Pro} = S^{in}_{Ac} = 880 \ g \ m^{-3}$, $S^{in}_{CH_4} = 0$; RUN4: $S^{in}_{Su} = 0$, $S^{in}_{Bu} = S^{in}_{Pro} = S^{in}_{Ac} = 1170 \ g \ m^{-3}$, $S^{in}_{CH_4} = 0$. $S^{in}_{Su}$: Sugar, $S^{in}_{Bu}$: Butyrate, $S^{in}_{Pro}$: Propionate, $S^{in}_{Ac}$: Acetate, $S^{in}_{CH_4}$: Methane}
          \label{f4.2.2} 
 \end{figure*}

\begin{figure*}
 \fbox{\includegraphics[width=1\textwidth, keepaspectratio]{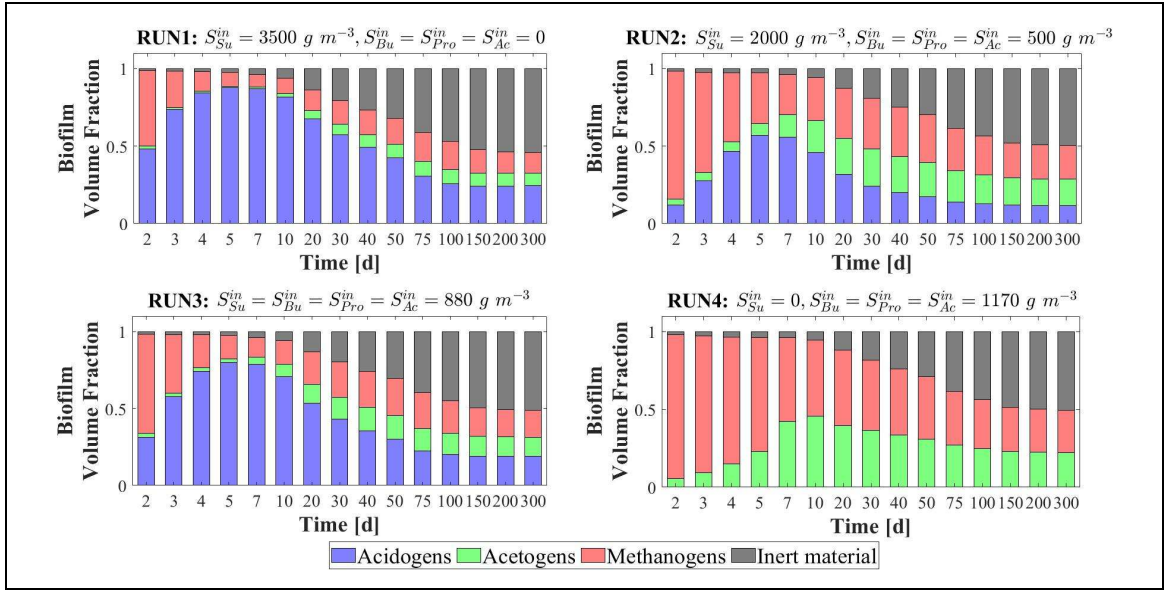}}   
 \caption{NS2 - Relative abundances of microbial species within the granule at several times under different influent wastewater compositions. $S^{in}_{Su}$: Sugar, $S^{in}_{Bu}$: Butyrate, $S^{in}_{Pro}$: Propionate, $S^{in}_{Ac}$: Acetate, $S^{in}_{CH_4}$: Methane}
          \label{f4.2.3} 
 \end{figure*}

\begin{figure*}
 \fbox{\includegraphics[width=1\textwidth, keepaspectratio]{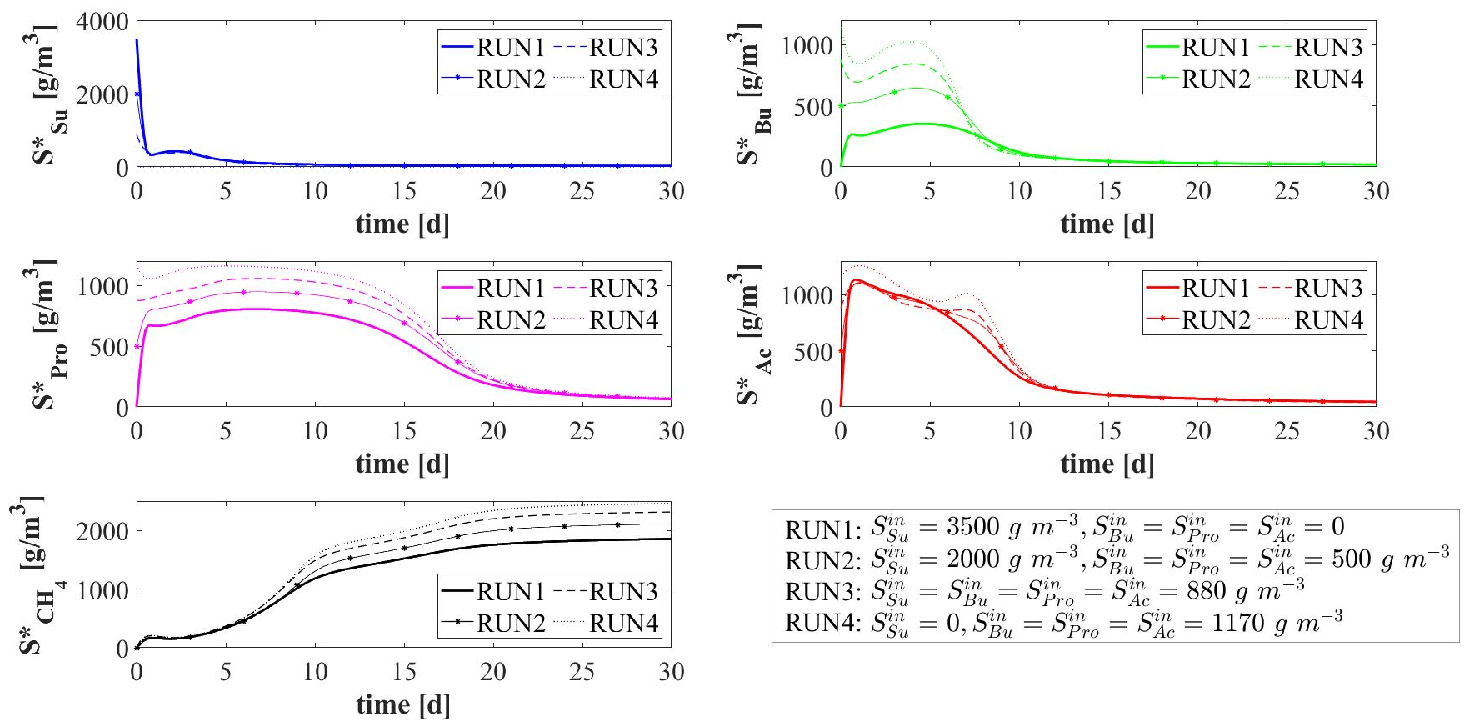}}   
 \caption{NS2 - Evolution of soluble substrates concentrations within the bulk liquid for different influent wastewater compositions.  $S^*_{Su}$: Sugar, $S^*_{Bu}$: Butyrate, $S^*_{Pro}$: Propionate, $S^*_{Ac}$: Acetate, $S^*_{CH_4}$: Methane}
          \label{f4.2.4} 
 \end{figure*}

\begin{figure*}
 \fbox{\includegraphics[width=1\textwidth, keepaspectratio]{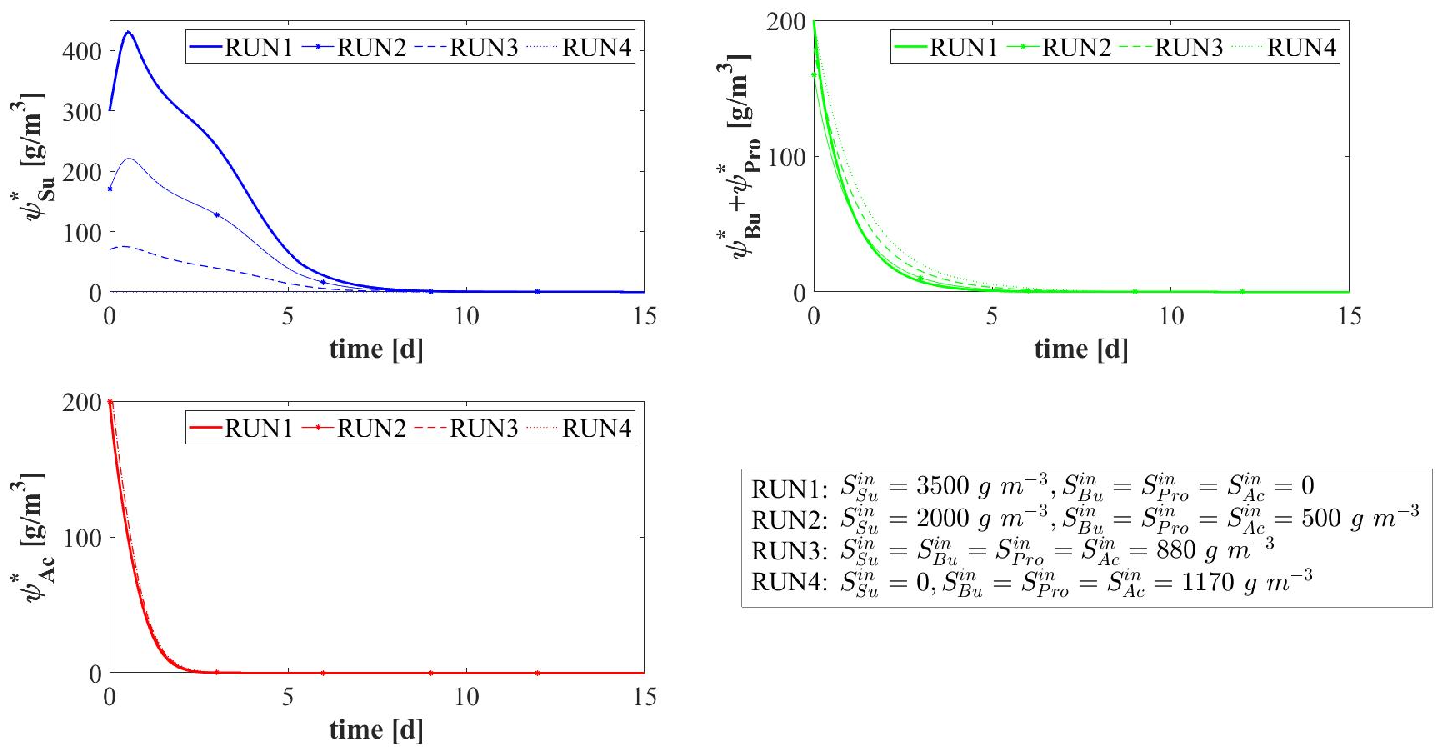}}   
 \caption{NS2 - Evolution of planktonic biomass concentrations within the bulk liquid for different influent wastewater compositions. $\psi^*_{Su}$: Sugar fermenters, $\psi^*_{Bu}$: Butyrate consumers, $\psi^*_{Pro}$: Propionate consumers, $\psi^*_{Ac}$: Acetoclastic methanogens}
          \label{f4.2.5} 
 \end{figure*}

The results presented in the previous section describe the dynamic evolution and the steady-state configuration of anaerobic granular biofilms growing in a sugar-fed reactor. However, the composition of the influent wastewater affects the granulation process and regulates the ecological succession and the growth of the individual species. Since the influent wastewater treated in anaerobic granular systems originate from various applications and present variable compositions of the organic load, it is interesting to compare the model results for different types of influent wastewater. In particular, in this study (NS2) different influent compositions and reactor inocula are set as model input: (RUN2: $S^{in}_{Su}=2000 \ g \ m^{-3}$, $S^{in}_{Bu}=S^{in}_{Pro}=S^{in}_{Ac}=500 \ g \ m^{-3}$, $S^{in}_{CH_4}=0$, $\psi^*_{Su,0} = 170 \ g \  m^{-3}$, $\psi^*_{Bu,0} = \psi^*_{Pro,0} = 40 \ g \  m^{-3}$, $\psi^*_{Ac,0} = 100 \ g \  m^{-3} $; RUN3: $S^{in}_{Su}=S^{in}_{Bu}=S^{in}_{Pro}=S^{in}_{Ac}=880 \ g \ m^{-3}$, $S^{in}_{CH_4}=0$, $\psi^*_{Su,0} = 70 \ g \  m^{-3}$, $\psi^*_{Bu,0} = 50 \ g \  m^{-3}, \psi^*_{Pro,0} = 40 \ g \  m^{-3}$, $\psi^*_{Ac,0} = 110 \ g \  m^{-3} $; RUN4: $S^{in}_{Su}=S^{in}_{CH_4}=0$, $S^{in}_{Bu}=S^{in}_{Pro}=S^{in}_{Ac}=1170 \ g \ m^{-3}$, $\psi^*_{Su,0} = 0$, $\psi^*_{Bu,0} = 60 \ g \  m^{-3}, \psi^*_{Pro,0} = 40 \ g \  m^{-3}$, $\psi^*_{Ac,0} = 110 \ g \  m^{-3} $). These cases have been compared with the case of reactor fed with only sugar (RUN1). The results are summarized in Figs. \ref{f4.2.1}-\ref{f4.2.5}.

Fig. \ref{f4.2.1} shows the trend of the granule radius $R(t)$ over time. Granules of different sizes are formed. These differences are related to the sessile biomass growth which varies according to the substrates present in the influent wastewater. Specifically, sessile growth is affected by anabolic and catabolic pathways of the microbial metabolism: the yield of acidogens on sugar $Y_{Su}$ is higher than the yields of the other species, thus, the amount of acidogenic biomass grown per unit of substrate consumed is higher than the other species. Therefore, the steady-state dimension of the granule increases with increasing sugar concentration in the influent $S^{in}_{Su}$. For $S^{in}_{Su}=0$ (RUN4), the granule achieves the smallest dimension.

The distribution of sessile biomass within the granule in the four cases is reported in Fig. \ref{f4.2.2}, at $T = 70 \ d$. As the sugar concentration in the influent $S^{in}_{Su}$ increases, the fraction of acidogens $f_{Su}$ (blue) increases, especially in the outer part of the granule, where there is maximum availability of substrate. When the sugar concentration $S^{in}_{Su}$ decreases from $3500 \ g \ m^{-3}$ (RUN1) to $880 \ g \ m^{-3}$ (RUN3), the acidogenic fraction present in the external part of the biofilm significantly reduces. Obviously, no acidogens are found within the granule when sugar is absent in the influent wastewater (RUN4). In addition, as the concentration of butyrate $S^{in}_{Bu}$, propionate $S^{in}_{Pro}$ and acetate $S^{in}_{Ac}$ in the influent increases going from RUN1 to RUN4, an increase in the fractions of acetogens $f_{Bu}+f_{Pro}$ (green) and methanogens $f_{Ac}$ (red) is observed.

The relative abundance of sessile microbial species is reported for different simulation times in Fig. \ref{f4.2.3}. When sugar is present in the influent (RUN1, RUN2, RUN3), the initial phase of the granulation is governed by acidogens (which have a higher growth rate) and methanogens (which have high attachment velocities). The acidogenic fraction (blue) reaches a maximum after $7\ d$ and then decreases when the availability of sugar in the bulk liquid reduces. When sugar is not present in the influent (RUN4), the granulation process is dominated by methanogens (red) and acetogens in small amounts (green). In all four cases, the maximum fraction of methanogens is observed at the beginning of the process due to their granulation properties. Then, the methanogenic fraction reduces due to the decay process and the competition with acidogens and acetogens. Furthermore, the acetogenic fraction is negligible in all cases during the initial phase of the granulation and grows when other microbial species become less competitive, and sugar is converted to butyrate and propionate. The microbial relative abundances related to the steady-state value confirm the results introduced in Fig. \ref{f4.2.2}: the fraction of acidogens increases with the increase of the sugar in the influent; the fractions of methanogens and acetogens increase with increasing VFAs in the influent; in all cases, inactive biomass (black) represents approximately 50\% of the total sessile biomass within the granule.
 
The model results related to the bulk liquid are summarized in Fig. \ref{f4.2.4} and Fig. \ref{f4.2.5}, which show how the concentration of soluble substrates and planktonic biomass changes over time. As reported in Fig. \ref{f4.2.4}, the composition of the influent wastewater affects the trend of the substrates mostly in the initial phase. In all cases, the AD process is completed in about  $30$ days: sugar (blue), butyrate (green), propionate (magenta) and acetate (red) are totally consumed and the concentration of methane (black) achieves a steady-state value. Notably, different productions of methane are observed as the composition of the influent changes. Concerning the concentration of planktonic biomass shown in Fig. \ref{f4.2.5}, the concentration of acidogens (blue) is affected by the composition of the influent in the initial phase of the granulation process ($0$-$7$ days). Conversely, the concentration of acetogens (green) and methanogens (red) in planktonic form have low growth rates and are washed out due to the dilution phenomenon even when significant concentrations of VFAs are present in the influent.

\subsection{NS3 - Effects of granulation properties} 
\label{n4.3}

The characteristics of the microbial community play a fundamental role in the anaerobic granulation process. In particular, such process is highly affected by the granulation properties of the planktonic biomass present in the reactor. The granulation properties of microbial cells depend on several factors such as hydrodynamic conditions inside the reactor, cell dimension, cell structure, ability to produce EPS and ability to use quorum sensing strategies. In this regard, several studies report that quorum sensing regulates the formation of filamentous cells frequently involved in anaerobic granular reactors, such as \textit{Methanosaeta} \cite{zhang2012acyl,li2014characterization}. As reported in \cite{li2015significant}, some quorum sensing molecule-promoted filamentous cells of \textit{Methanosaeta harundinacea} enhance the reactor performance, improve the granulation, and immobilize other synergistically functioning bacterial groups.  

In the management of full-scale reactors, several strategies are pursued to reduce the duration of time of granule formation and improve the efficiency of the wastewater treatment process. For example, the direct addition of the quorum sensing molecule acyl homoserine lactone (AHL) during granule formation might remarkably improve the granulation process in anaerobic granular reactors \cite{li2015significant,zhang2012acyl,de2016perspectives}. Furthermore, bioaugmentation is regarded as a promising method to improve the granulation velocity and reduce the time required for the start-up of full-scale plants \cite{guiot2000immobilization,nancharaiah2008bioaugmentation,jin2014bio}. It consists in the addition of selected microbial cultures with high self-aggregation ability within the reactor \cite{bathe2005aerobic}. Overall, such strategies positively alter the granulation properties of the planktonic microbial community.

In this framework, a numerical study (NS3) is conducted to investigate the effects of biomass granulation properties on the process of granule formation. For this purpose, nine simulations (RUN5 - RUN13) are carried out with different attachment velocities $v_{a,i}$. The nine values of $v_{a,i}$ used are chosen by increasing and reducing the default values (presented in Table \ref{t3.1}) through different multiplication factors ($0.05, 0.1, 0.25, 0.5, 1, 2, 3, 4, 5$). The concentration of soluble substrates in the influent wastewater $S^{in}_j$ and the initial concentration of planktonic biomasses within the reactor $\psi^*_{i,0}$ set for this numerical study are the same used in NS1 and are reported in Table \ref{t4.1}.

 \begin{figure*}
 \fbox{\includegraphics[width=1\textwidth, keepaspectratio]{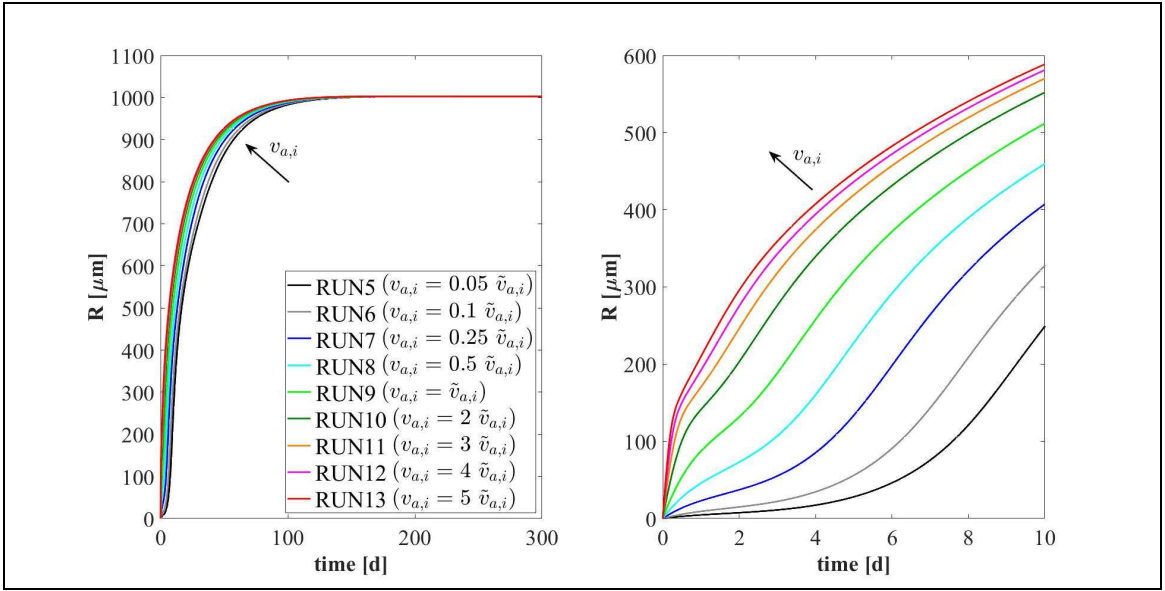}}   
 \caption{NS3 - Biofilm radius evolution over time for different attachment velocities $v_{a,i}$ (left), with a focus on the first 10 days (right). $\tilde{v}_{a,i}$: value of attachment velocity of the $i^{th}$ planktonic species set in RUN1. Influent wastewater composition: $S^{in}_{Su} = 3500 \ g \ m^{-3}$ (Sugar), $S^{in}_{Bu} = 0$ (Butyrate), $S^{in}_{Pro} = 0$ (Propionate), $S^{in}_{Ac} = 0$ (Acetate), $S^{in}_{CH_4} = 0$ (Methane)}
          \label{f4.3.1} 
 \end{figure*}

\begin{figure*}
 \fbox{\includegraphics[width=1\textwidth, keepaspectratio]{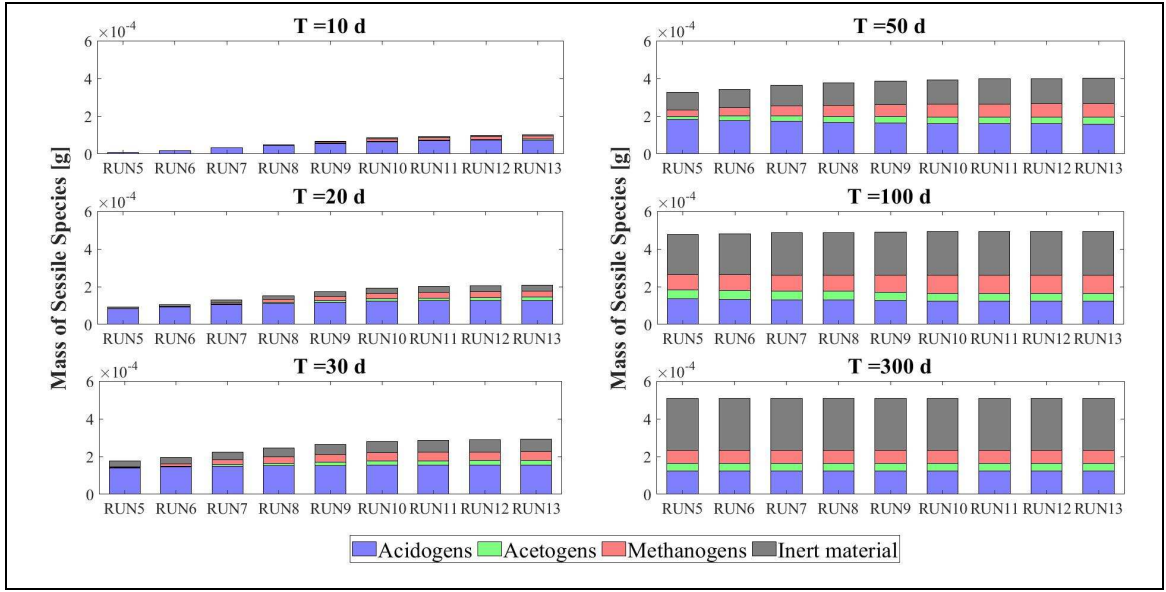}}   
 \caption{NS3 - Mass of microbial species within the granule at $T = 10 \ d$, $T = 20 \ d$, $T = 30 \ d$, $T = 50 \ d$, $T = 100 \ d$ and $T = 300 \ d$ under different attachment velocities $v_{a,i}$. RUN5: $v_{a,i} = 0.05 \ \tilde{v}_{a,i}$, RUN6: $v_{a,i} = 0.1 \ \tilde{v}_{a,i}$, RUN7: $v_{a,i} = 0.25 \ \tilde{v}_{a,i}$, RUN8: $v_{a,i} = 0.5 \ \tilde{v}_{a,i}$, RUN9: $v_{a,i} =  \ \tilde{v}_{a,i}$, RUN10: $v_{a,i} = 2 \ \tilde{v}_{a,i}$, RUN11: $v_{a,i} = 3 \ \tilde{v}_{a,i}$, RUN12: $v_{a,i} = 4 \ \tilde{v}_{a,i}$, RUN13: $v_{a,i} = 5 \ \tilde{v}_{a,i}$. $\tilde{v}_{a,i}$: value of attachment velocity of the $i^{th}$ planktonic species set in RUN1. Influent wastewater composition: $S^{in}_{Su} = 3500 \ g \ m^{-3}$ (Sugar), $S^{in}_{Bu} = 0$ (Butyrate), $S^{in}_{Pro} = 0$ (Propionate), $S^{in}_{Ac} = 0$ (Acetate), $S^{in}_{CH_4} = 0$ (Methane)}
          \label{f4.3.2} 
 \end{figure*}

\begin{figure*}
 \fbox{\includegraphics[width=1\textwidth, keepaspectratio]{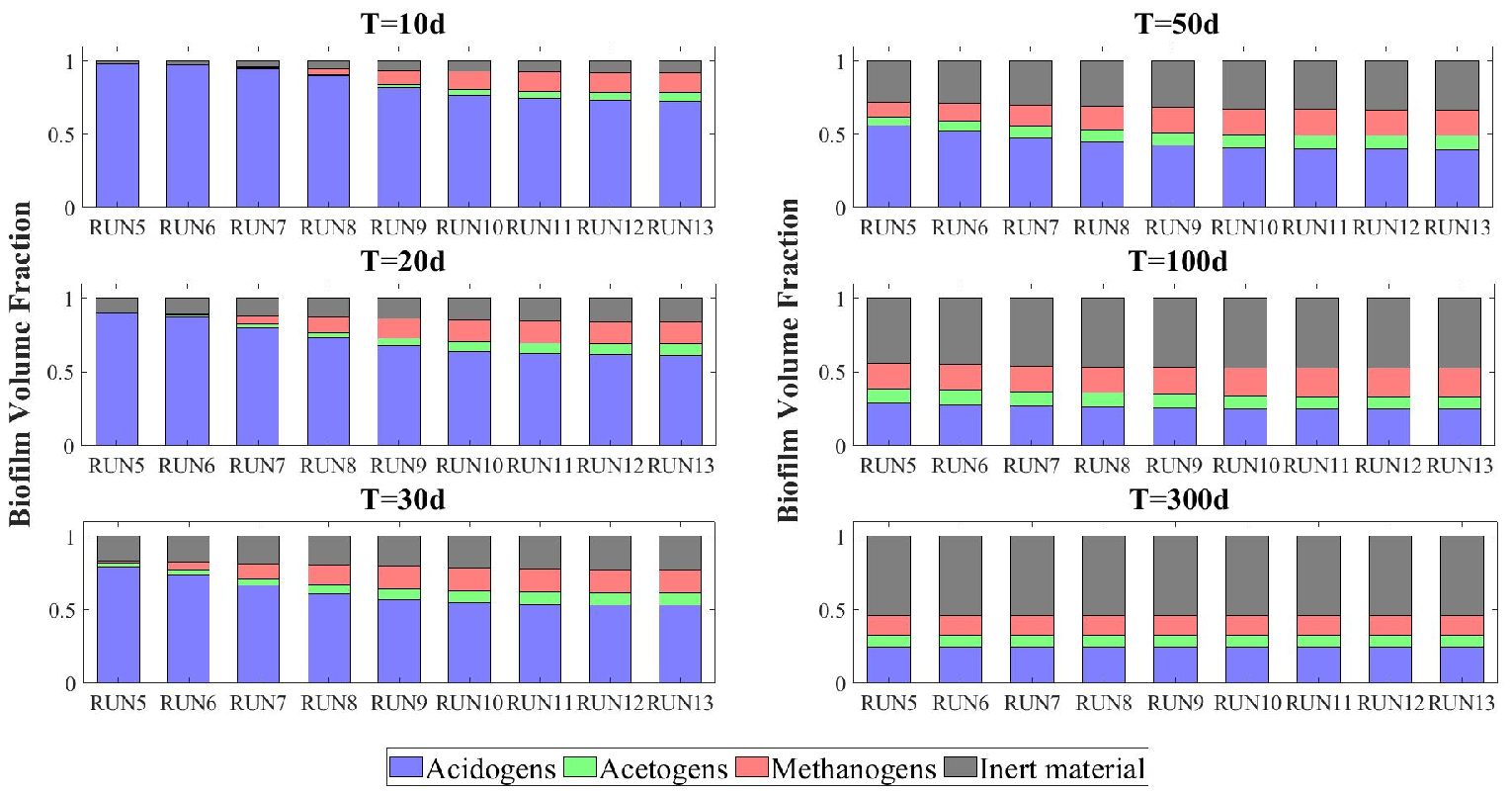}}   
 \caption{NS3 - Relative abundances of microbial species within the granule at $T = 10 \ d$, $T = 20 \ d$, $T = 30 \ d$, $T = 50 \ d$, $T = 100 \ d$ and $T = 300 \ d$ under different attachment velocities $v_{a,i}$. RUN5: $v_{a,i} = 0.05 \ \tilde{v}_{a,i}$, RUN6: $v_{a,i} = 0.1 \ \tilde{v}_{a,i}$, RUN7: $v_{a,i} = 0.25 \ \tilde{v}_{a,i}$, RUN8: $v_{a,i} = 0.5 \ \tilde{v}_{a,i}$, RUN9: $v_{a,i} =  \ \tilde{v}_{a,i}$, RUN10: $v_{a,i} = 2 \ \tilde{v}_{a,i}$, RUN11: $v_{a,i} = 3 \ \tilde{v}_{a,i}$, RUN12: $v_{a,i} = 4 \ \tilde{v}_{a,i}$, RUN13: $v_{a,i} = 5 \ \tilde{v}_{a,i}$. $\tilde{v}_{a,i}$: value of attachment velocity of the $i^{th}$ planktonic species set in RUN1. Influent wastewater composition: $S^{in}_{Su} = 3500 \ g \ m^{-3}$ (Sugar), $S^{in}_{Bu} = 0$ (Butyrate), $S^{in}_{Pro} = 0$ (Propionate), $S^{in}_{Ac} = 0$ (Acetate), $S^{in}_{CH_4} = 0$ (Methane)}
          \label{f4.3.3} 
 \end{figure*}

\begin{figure*}
 \fbox{\includegraphics[width=1\textwidth, keepaspectratio]{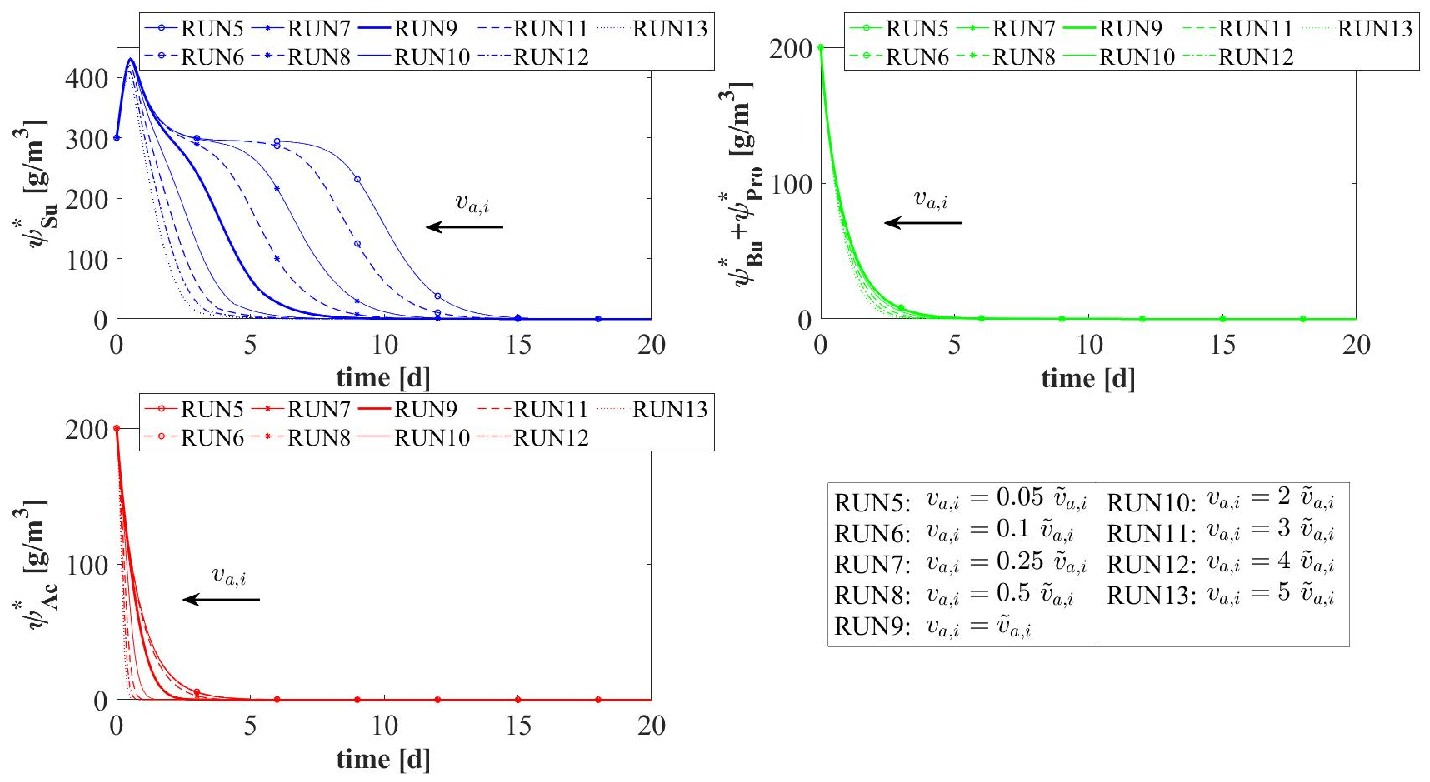}}   
 \caption{NS3 - Evolution of planktonic biomass concentrations within the bulk liquid for different attachment velocities $v_{a,i}$. $\psi^*_{Su}$: Sugar fermenters, $\psi^*_{Bu}$: Butyrate consumers, $\psi^*_{Pro}$: Propionate consumers, $\psi^*_{Ac}$: Acetoclastic methanogens.  $\tilde{v}_{a,i}$: value of attachment velocity of the $i^{th}$ planktonic species set in RUN1. Influent wastewater composition: $S^{in}_{Su} = 3500 \ g \ m^{-3}$ (Sugar), $S^{in}_{Bu} = 0$ (Butyrate), $S^{in}_{Pro} = 0$ (Propionate), $S^{in}_{Ac} = 0$ (Acetate), $S^{in}_{CH_4} = 0$ (Methane)}
          \label{f4.3.4} 
 \end{figure*}

\begin{figure*}
 \fbox{\includegraphics[width=1\textwidth, keepaspectratio]{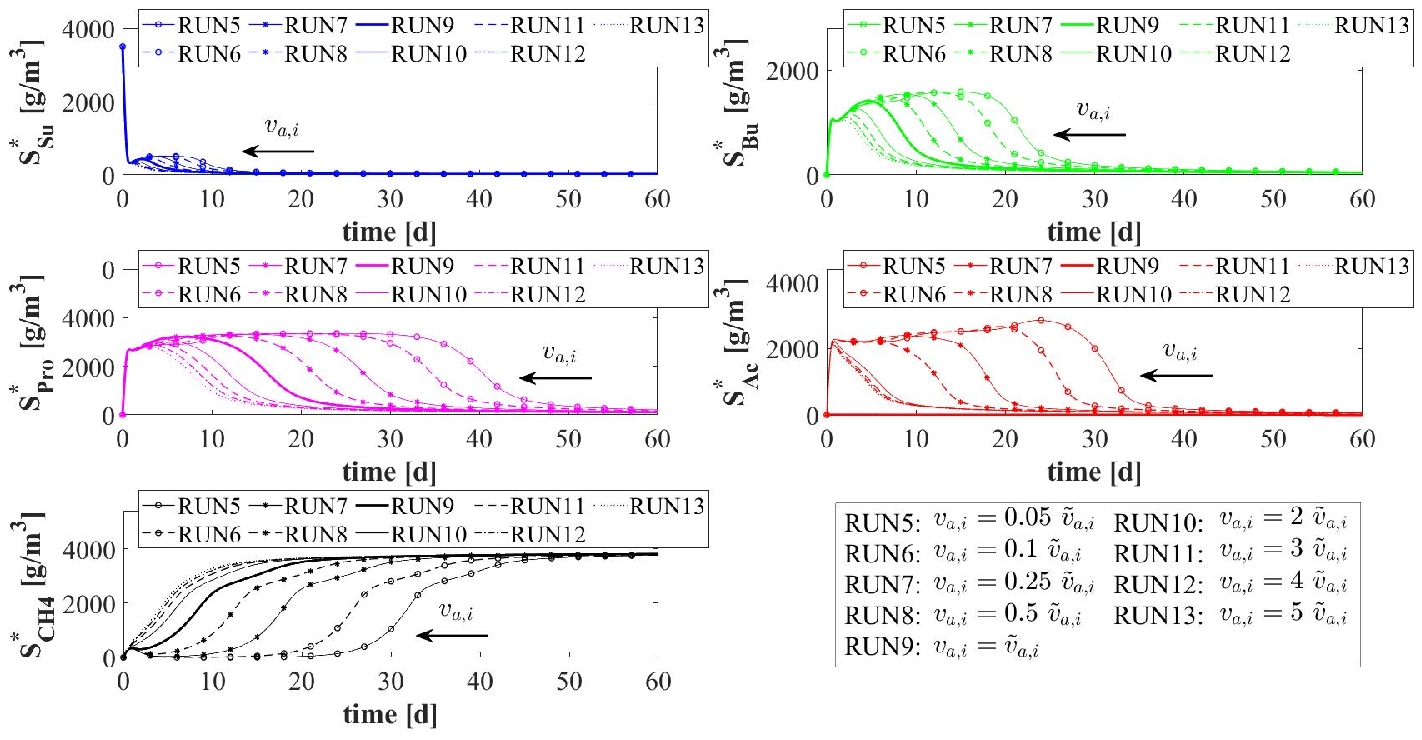}}   
 \caption{NS3 - Evolution of soluble substrates concentrations within the bulk liquid for different attachment velocities $v_{a,i}$. $S^*_{Su}$: Sugar, $S^*_{Bu}$: Butyrate, $S^*_{Pro}$: Propionate, $S^*_{Ac}$: Acetate, $S^*_{CH_4}$: Methane. $\tilde{v}_{a,i}$: value of attachment velocity of the $i^{th}$ planktonic species set in RUN1. Influent wastewater composition: $S^{in}_{Su} = 3500 \ g \ m^{-3}$ (Sugar), $S^{in}_{Bu} = 0$ (Butyrate), $S^{in}_{Pro} = 0$ (Propionate), $S^{in}_{Ac} = 0$ (Acetate), $S^{in}_{CH_4} = 0$ (Methane)}
          \label{f4.3.5} 
 \end{figure*}

The results of this study are reported in Figs. \ref{f4.3.1}-\ref{f4.3.5}. The evolution of the granule radius $R(t)$ over time is shown in Fig. \ref{f4.3.1}. From Fig. \ref{f4.3.1} (right) it is clear that different attachment velocities $v_{a,i}$ lead to different growth rates of the granule in the initial phase of the process: when the inoculated microbial community is more inclined to grow in sessile form, the granulation process occurs faster and the granule reaches earlier the steady-state dimension. However, such steady-state dimension is not dependent on the attachment velocity. Indeed, the profiles of $R(t)$ for different $v_{a,i}$ get closer over time and reach the same steady-state value. 
 
Fig. \ref{f4.3.2} and Fig. \ref{f4.3.3} report the mass and the relative abundance of the different sessile microbial species within the granule, respectively. Again, relevant differences concern the initial phase ($ T = 10$-$20 \ d $), when the total sessile mass, proportional to the granule dimension, is higher in the case of more intense attachment process. However, after long times both the total sessile mass and the relative abundance of individual microbial species within the granule are no longer affected by $v_{a,i}$ and all simulations achieve the same steady-state configuration.
 
Other interesting results refer to the effects that the granulation process has on the planktonic biomass $\psi^*_{i}$ (Fig. \ref{f4.3.4}) and soluble substrates $S^*_{j}$ (Fig. \ref{f4.3.5}) within the reactor. The concentration profiles of the planktonic acetogens $\psi^*_{Bu}+\psi^*_{Pro}$ (green) and methanogens $\psi^*_{Ac}$ (red) in the bulk liquid shown in Fig. \ref{f4.3.4} are not very sensitive to the variation of $v_{a,i}$. Indeed, the reduction of the planktonic biomass $\psi^*_{i}$ depends on two phenomena: attachment and dilution. The reduction of $\psi^*_{Bu}, \psi^*_{Pro}$ and $\psi^*_{Ac}$ occurs in the initial phase of the process, when the granules are small and the attachment flux of planktonic biomass (proportional to the granule surface $A(t)$) has limited effects on the properties of the bulk liquid. On the other hand, the dilution process is prominent: the hydrodynamic conditions (high flow rate, low HRT) and the slow metabolic growth (due to low maximum growth rates and substrate unavailability) lead to the washout of planktonic acetogens and methanogens. Such dilution process is not affected by granulation properties and therefore leads to similar profiles by varying $v_{a,i}$. Conversely, the planktonic acidogens $\psi^*_{Su}$ (blue) have higher growth rates and optimal conditions to grow (sugar-rich influent), hence, they populate the reactor for longer times and decrease mainly due to the granulation process, which is strongly influenced by $v_{a,i}$. Consequently, different values of $v_{a,i}$ correspond to different profiles of planktonic acidogens $\psi^*_{Su}$: the higher $v_{a,i}$, the faster the reduction of the concentration $\psi^*_{Su}$ in the bulk liquid.

Fig. \ref{f4.3.5} shows the trend of soluble substrates within the reactor. As just mentioned, in the first few days the granules have small dimensions and the consumption and production of soluble substrates mainly depend on the metabolic activity of the planktonic biomass. Consequently, in the initial phase the trends of soluble substrates are not affected by attachment conditions. For later times, the granule size and the amount of sessile biomass within the reactor increase and the trend of the substrates becomes more sensitive to $v_{a,i}$: for small values of $v_{a,i}$ the concentrations of soluble substrates reach steady-state conditions in $40$-$50$ days while for high values of $v_{a,i}$ they require half of the time to reach a steady-state configuration. 

\subsection{NS4 - Effects of biomass density} 
\label{n4.4}

Biomass density of the granules involved in granular biofilm systems is highly variable due to several factors, such as hydrodynamic conditions, shear forces, EPS production. Firstly, high shear forces lead to stronger and denser granules, while weaker and more porous granule structures develop for lower shear forces \cite{liu2002essential,di2006influence,tay2001effects}. Moreover, EPS production is generally thought to increase cell surface hydrophobicity and promote the formation of a sticky matrix which favours the granulation of new cells or flocs \cite{trego2019granular}. Thus, EPS positively influences the granulation process, contributing to maintain the structural integrity of the biofilm matrix and improve the biomass density. Biomass density is a crucial property of granular biofilms because it regulates the mass transfer, the consumption of soluble substrates within the granules and consequently the growth of microbial species and granule dynamics. As a result, granules of different densities typically have different dimensions and are characterized by different microbial stratifications.

\begin{figure*}
 \fbox{\includegraphics[width=1\textwidth, keepaspectratio]{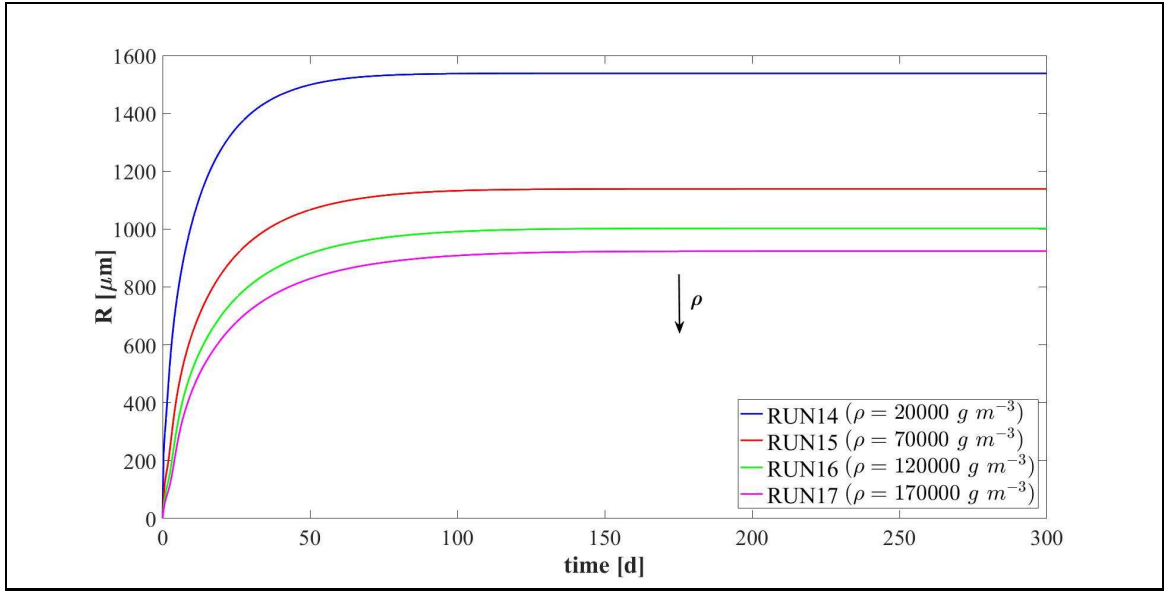}}   
 \caption{NS4 - Biofilm radius evolution over time for different biomass densities $\rho$. Influent wastewater composition: $S^{in}_{Su} = 3500 \ g \ m^{-3}$ (Sugar), $S^{in}_{Bu} = 0$ (Butyrate), $S^{in}_{Pro} = 0$ (Propionate), $S^{in}_{Ac} = 0$ (Acetate), $S^{in}_{CH_4} = 0$ (Methane)}
          \label{f4.4.1} 
 \end{figure*}

\begin{figure*}
 \fbox{\includegraphics[width=1\textwidth, keepaspectratio]{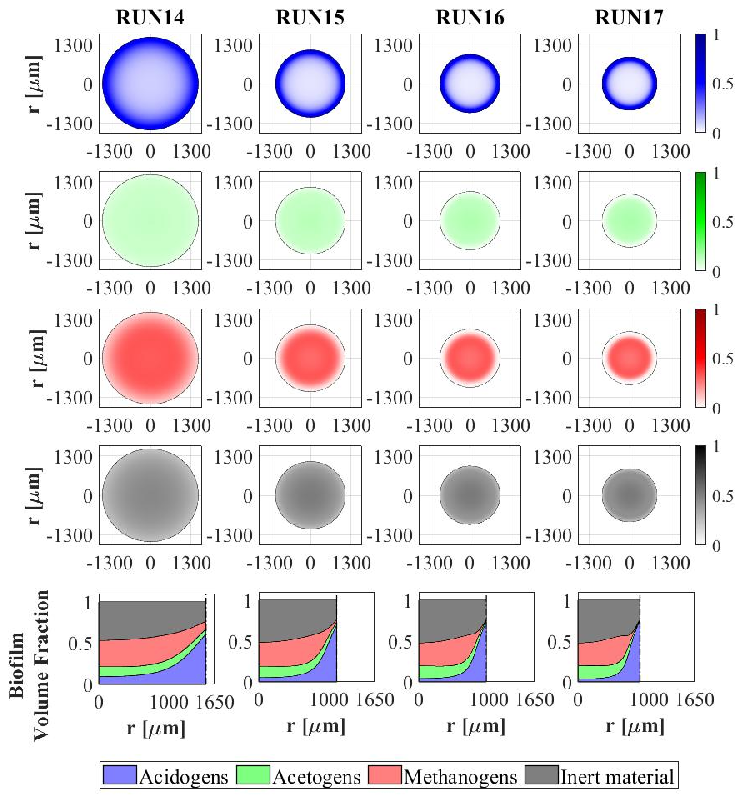}}   
 \caption{NS4 - Microbial species distribution in the diametrical section and across the radius of the granule at $T = 70 \ d$, for different biomass densities. RUN14: $\rho = 20000 \ g \ m^{-3}$, RUN15: $\rho = 70000 \ g \ m^{-3}$, RUN16: $\rho = 120000 \ g \ m^{-3}$, RUN17: $\rho = 170000 \ g \ m^{-3}$. Influent wastewater composition: $S^{in}_{Su} = 3500 \ g \ m^{-3}$ (Sugar), $S^{in}_{Bu} = 0$ (Butyrate), $S^{in}_{Pro} = 0$ (Propionate), $S^{in}_{Ac} = 0$ (Acetate), $S^{in}_{CH_4} = 0$ (Methane)}
          \label{f4.4.2} 
 \end{figure*}

\begin{figure*}
 \fbox{\includegraphics[width=1\textwidth, keepaspectratio]{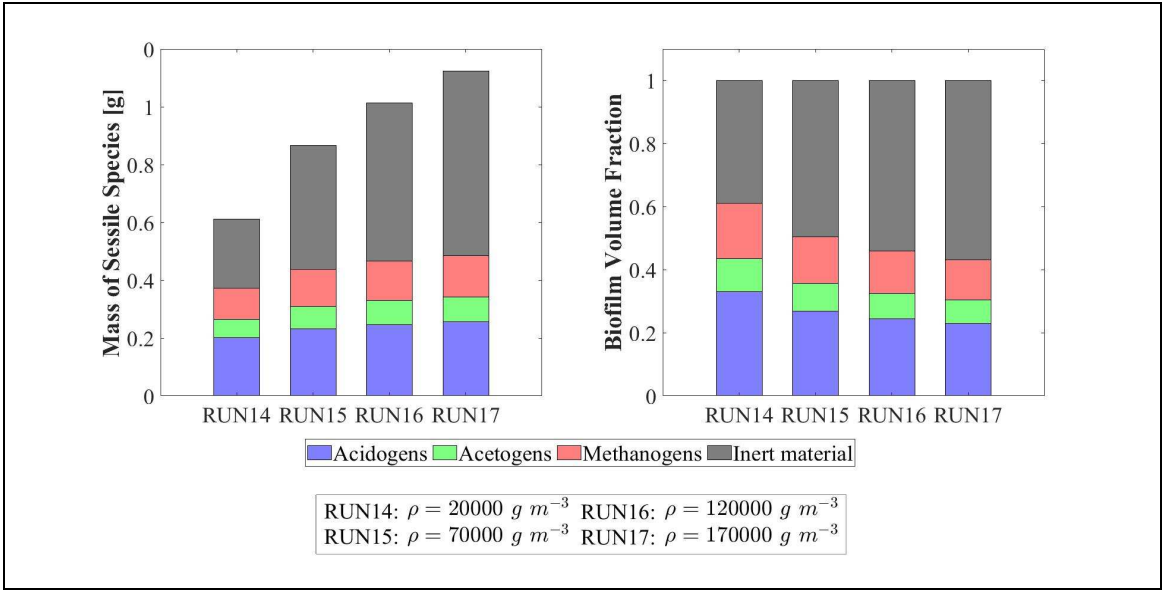}}   
 \caption{NS4 - Mass (left) and relative abundances (right) of microbial species within the granule at $T = 300 \ d$ for different biomass densities $\rho$. Influent wastewater composition: $S^{in}_{Su} = 3500 \ g \ m^{-3}$ (Sugar), $S^{in}_{Bu} = 0$ (Butyrate), $S^{in}_{Pro} = 0$ (Propionate), $S^{in}_{Ac} = 0$ (Acetate), $S^{in}_{CH_4} = 0$ (Methane)}
          \label{f4.4.3} 
 \end{figure*}

In this context, a numerical study (NS4) is performed to describe the evolution of biofilm granules with different biomass densities. Four simulations (RUN14 - RUN17) are carried out using four different values of biomass density $\rho$ (RUN14: $\rho = 20000 \ g \  m^{-3}$, RUN15: $\rho = 70000 \ g \  m^{-3}$, RUN16: $\rho = 120000 \ g \  m^{-3}$, RUN17: $\rho = 170000 \ g \  m^{-3}$). The concentration of soluble substrates in the influent wastewater $S^{in}_j$ and the initial concentration of planktonic biomasses within the reactor $\psi^*_{i,0}$ set for this numerical study are the same used in NS1 and are reported in Table \ref{t4.1}. Numerical results are summarized in Figs. \ref{f4.4.1}-\ref{f4.4.3}.

The evolution of the granule radius $R(t)$ over time is shown in Fig. \ref{f4.4.1}. It is clear that higher is the biomass density of the granule, smaller is the steady-state radius achieved. This is due to different mass transfer conditions occurring within the granule: higher biomass densities entail higher flux of soluble substrates exchanged between bulk liquid and granules. For this reason, for higher densities, the substrates in the bulk liquid are consumed faster and the metabolic growth rates driving the growth of the granule are on average lower. This leads to smaller granules, in accordance with \cite{liu2002essential}.

The microbial ecology for the four values of $\rho$ and at $ T = 70 \ d $ is reported in Fig. \ref{f4.4.2}. As can be seen, granules with different densities have different microbial distributions. In particular, when $\rho=20000 \ g \  m^{-3}$ (RUN14) an homogeneous distribution is observed: although acidogens (blue) have a tendency to gather in the outermost layers and methanogens (red) and acetogens (green) have the tendency to populate the internal part, the microbial distribution within the granule is fairly homogeneous. This is due to the mass transfer of the soluble substrates within the granule: a low biomass density leads to small gradients of soluble substrates across the granule and an homogeneous growth of the different microbial species is observed throughout the biofilm. As the density of the biofilm increases (RUN15-RUN17), the gradient of soluble substrates across the biofilm increases and the layered distribution of the biomass is clearer and visible: acidogens are strictly confined in the outer layer while acetogens and methanogens are mostly present in the inner part. These results are in agreement with the experimental evidence reported in \cite{batstone2004influence}.
  
In Fig. \ref{f4.4.3} the mass (left) and the relative abundance (right) of the sessile microbial species within the biofilm are shown at the steady-state condition ($T=300 \ d$). Higher biomass density leads higher amount (left) and fractions (right) of dead biomass accumulated as inert (black).

\subsection{NS5 - Effects of erosive detachment} 
\label{n4.5}

The evolution of the granule dimension is the result of a dynamic equilibrium between biomass growth, attachment of new biomass and detachment process. The detachment flux is essentially related to the erosion process occurring on the granule surface, due to the effect of the hydrodynamic shear forces developing in the reactor \cite{arcand1994impact}. These forces are highly variable due to the influence of several factors, such as liquid upflow velocity, gas production, particle-particle collision, eventual mixing systems and geometry of the reactor \cite{liu2002essential,tay2001effects}.

In this perspective, it is interesting to investigate the role of detachment phenomena induced by shear stress on the anaerobic granulation process and study its effects on the granule dimension and, consequently, on the distribution, amount and relative abundance of sessile biomass within the granule. For this purpose, the fifth and last study (NS5) is carried out based on 8 simulations (RUN18 - RUN25) with 8 different values of the detachment coefficient $\lambda$, in order to simulate different shear stress conditions. The values used are $\lambda = 4, 8, 12, 16, 20, 24, 28, 32 \ m^{-1} \ d^{-1}$. The concentration of soluble substrates in the influent wastewater $S^{in}_j$ and the initial concentration of planktonic biomasses within the reactor $\psi^*_{i,0}$ set for this numerical study are the same used in NS1 and are reported in Table \ref{t4.1}.

It should be noted that, contrary to the attachment, the detachment phenomenon increases as the biofilm dimension increases and has a negligible effect on the initial phase of the granulation process ($\sigma_d(t)=\lambda R^2(t)$). For this reason, the study does not focus on the initial biofilm formation but investigates the long-term effects of the detachment process.

  \begin{figure*}
 \fbox{\includegraphics[width=1\textwidth, keepaspectratio]{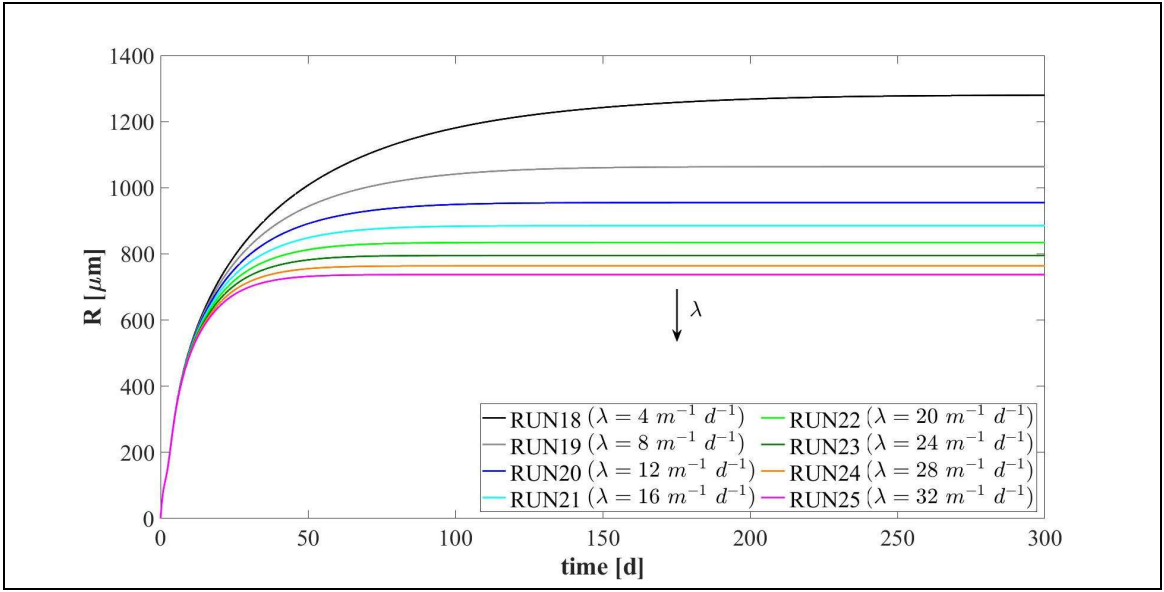}}   
 \caption{NS5 - Biofilm radius evolution over time for different detachment coefficients $\lambda$. Influent wastewater composition: $S^{in}_{Su} = 3500 \ g \ m^{-3}$ (Sugar), $S^{in}_{Bu} = 0$ (Butyrate), $S^{in}_{Pro} = 0$ (Propionate), $S^{in}_{Ac} = 0$ (Acetate), $S^{in}_{CH_4} = 0$ (Methane)}
          \label{f4.5.1} 
 \end{figure*}

Fig. \ref{f4.5.1} summarizes the effects that different detachment conditions have on the variation of the granule radius $R(t)$ over time and on the granule steady-state dimension. As explained above, the formation and the initial evolution of the granule are not affected by the detachment phenomenon. Indeed, it is clear that the trend of $R(t)$ is not influenced by $\lambda$ until $T = 10$-$20 \ d$. When the granule reaches a $600 \ \mu m$ radius, it becomes very sensitive to the detachment coefficient: as $\lambda$ increases, the erosion phenomenon increases, and a smaller steady-state granule dimension is achieved. However, the steady-state granule radius has a less than linear behaviour with increasing $\lambda$. Furthermore, in the case of positive attachment flux, steady-state $R(t)$ tends asymptotically to $0$ for $\lambda$ tending towards an infinite value. Indeed, when $R(t) = 0$ the detachment flux is null (see Eq.\ref{2.19}) and any positive value of attachment flux is enough to trigger the expansion of the spherical free boundary domain.

 \begin{figure*}
 \fbox{\includegraphics[width=1\textwidth, keepaspectratio]{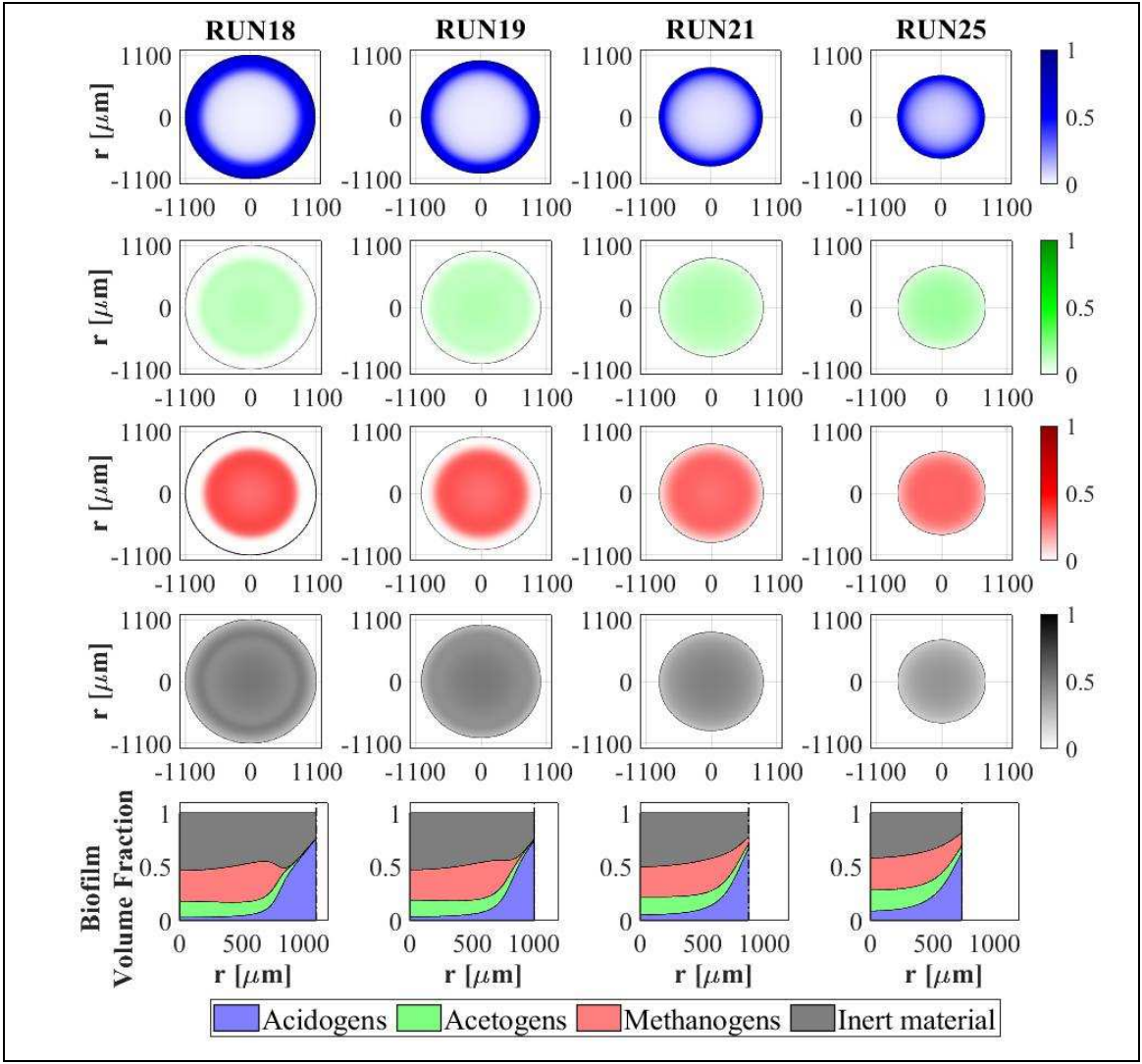}}   
 \caption{NS5 - Microbial species distribution in the diametrical section and across the radius of the granule at $T = 70 \ d$, for different detachment coefficients. RUN18: $\lambda = 4 \ m^{-1} \ d^{-1}$, RUN19: $\lambda = 8 \ m^{-1} \ d^{-1}$, RUN21: $\lambda = 16 \ m^{-1} \ d^{-1}$, RUN25: $\lambda = 32 \ m^{-1} \ d^{-1}$. Influent wastewater composition: $S^{in}_{Su} = 3500 \ g \ m^{-3}$ (Sugar), $S^{in}_{Bu} = 0$ (Butyrate), $S^{in}_{Pro} = 0$ (Propionate), $S^{in}_{Ac} = 0$ (Acetate), $S^{in}_{CH_4} = 0$ (Methane)}
          \label{f4.5.2} 
 \end{figure*}
 
 \begin{figure*}
 \fbox{\includegraphics[width=1\textwidth, keepaspectratio]{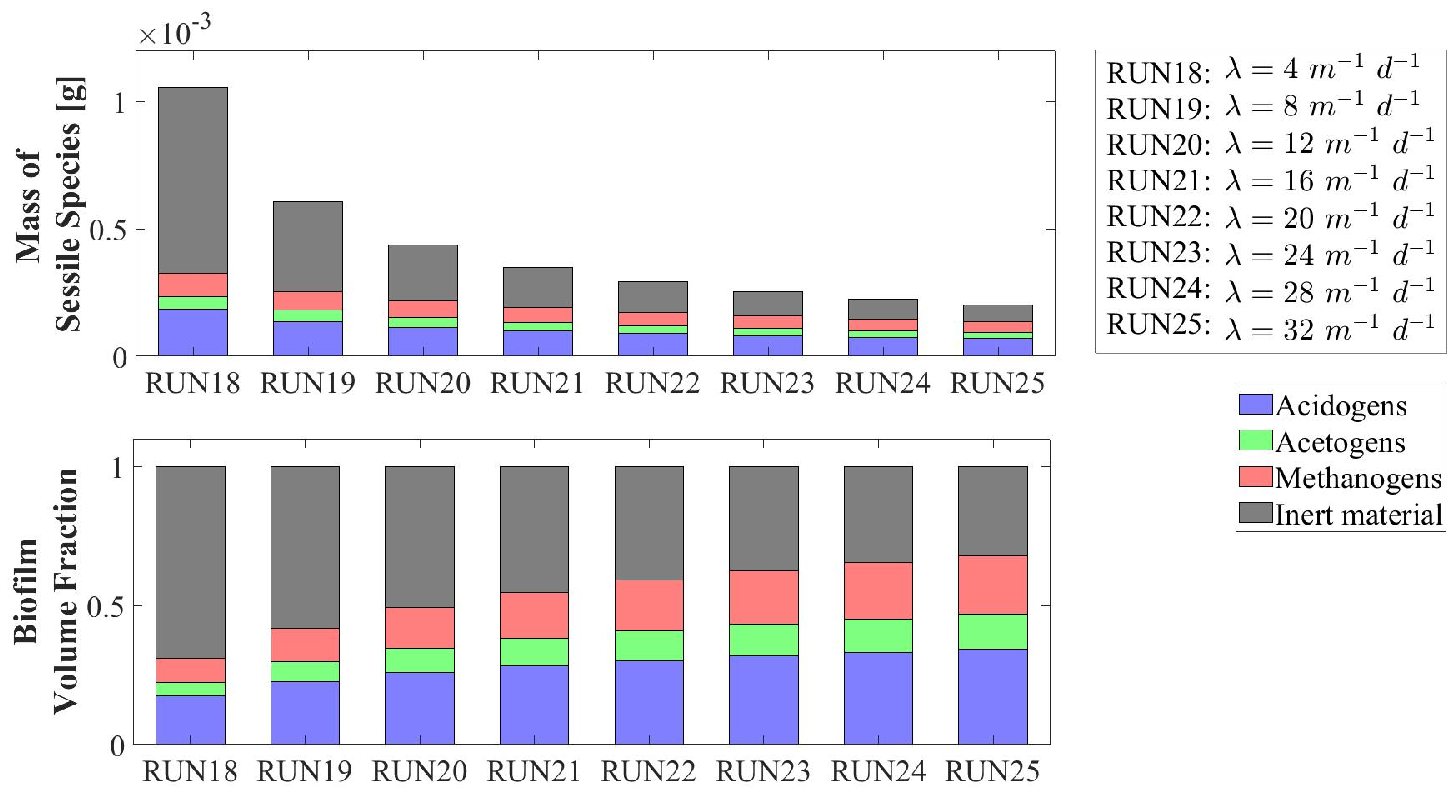}}   
 \caption{NS5  - Mass (top) and relative abundances (bottom) of microbial species within the granule at $T = 300 \ d$ for different detachment coefficients $\lambda$. Influent wastewater composition: $S^{in}_{Su} = 3500 \ g \ m^{-3}$ (Sugar), $S^{in}_{Bu} = 0$ (Butyrate), $S^{in}_{Pro} = 0$ (Propionate), $S^{in}_{Ac} = 0$ (Acetate), $S^{in}_{CH_4} = 0$ (Methane)}
          \label{f4.5.3} 
 \end{figure*}

Fig. \ref{f4.5.2} presents the distribution of sessile biomasses within the granule under four different detachment conditions ($\lambda = 4, 8, 16, 32 \ m^{-1} \ d^{-1}$), at $ T = 70 \ d $. As $\lambda$ increases, an increase in the active biomass fraction $f_{Su}+f_{Bu}+f_{Pro}+f_{Ac}$ and a reduction in the inert material $f_{I}$ occur within the granule. Moreover, for all $\lambda$ values the granule is mainly composed of acidogens (blue) in the external layers, and methanogens (red) and acetogens (green) in the internal nucleus. This distribution appears more evident when the granule is larger (low $\lambda $) and there are higher gradients of soluble substrates along the radius.

Finally, Fig. \ref{f4.5.3} reports the steady-state mass (top) and the relative abundance (bottom) of sessile microbial species within the granule under different detachment conditions, at $T=300 \ d$. Since the density is constant and equal in all simulations, the total sessile mass within the granule is directly proportional to the granule dimension. Then, a higher value of detachment coefficient $\lambda$ leads to a smaller granule and consequently, a smaller sessile mass both overall and for individual species (Fig. \ref{f4.5.3}-top). The relative abundances shown in Fig. \ref{f4.5.3} (bottom) confirm the results presented in Fig. \ref{f4.5.2}: higher erosion conditions lead to smaller granules which are characterized by a higher fraction of active biomass.

\section{Discussion} 
\label{n5}

\subsection{Model assumptions} 
\label{n5.1}

An anaerobic granular biofilm reactor is an extremely complex and heterogeneous multiphase biological system characterized by properties that vary over time and space. This system is constituted by a liquid medium where an assorted microbial community is immersed. Most of this community is organized into granular aggregates: spherical-shaped biological structures with variable densities composed of several microbial species in sessile form which enhance the spatial heterogeneity. The life cycle of the \textit{de novo} granules includes an initial phase of granulation which leads to the granule formation, a maturation phase in which the granule size increases and a final phase of breaking. The pieces of biofilm deriving from the breaking of a granule may in turn originate new granules. During this evolution, the granule is affected by complex phenomena which radically influence its structure and suddenly change its properties: granulation processes of planktonic biomass, metabolic growth and decay of sessile biomass, particle-particle interactions, EPS secretion, gas production, invasion processes by planktonic cells, detachment processes induced by intense hydrodynamic conditions and shear stress. The combination of these factors leads to a biological community consisting of a number of granules that vary over time, which differ from one another in size, density and microbial distribution. In addition, the location inside the reactor influences the characteristics of the granules. Indeed, due to the geometry and the hydrodynamic conditions of these systems, gradients in the concentrations of soluble substrates establish along the reactor and amplify the differences between the granules located at different points of the system. However, in some cases mixing devices are added to the system, in order to enhance the movement of the sludge and to reduce the gradients without increasing the flow rate and velocity.

Given its physical and biological complexity, the mathematical modelling of an anaerobic granular-based system necessarily requires the introduction of some model assumptions. In this perspective, the model presented in this work describes the anaerobic granular system as a domain having a constant liquid volume where a fixed number of granules is immersed. All granules are assumed to have identical evolution and properties (same size, same density, same constituents). The single granule is modelled as a spherical free boundary domain having a biomass density constant in time and space. Under the assumption of perfect mixing, the properties of the bulk liquid, in particular the concentration of soluble substrates and planktonic biomass, are supposed to be the same at every point and vary only over time. The attachment is modelled as a continuous, deterministic process which depends linearly on the concentration of the planktonic biomass. The detachment process is modelled as a continuous mass flux that detaches from the granule due to the effect of shear-induced erosion. Due to the uncertainty about the form of detached biomass, its contribution to planktonic and sessile biomass is neglected. Moreover, since the mechanisms leading to the breaking of the granule are not perfectly clear, breaking phenomena are not included in the model. The process of invasion by planktonic cells present in the bulk liquid is modelled as a diffusive transport across the granule. This eliminates the restrictions on the granule ecological structure that could be generated in particular cases, and guarantees the growth of each sessile microbial species where optimal conditions for its metabolism occur. Finally, suspended substrates in the influent wastewater, gas transfer processes and EPS production are neglected as they are not significant for the purposes of the numerical investigation presented in this work: qualitatively studying the anaerobic granulation process, the evolution of the granules over time and the related ecological succession. 

It is emphasized that all assumptions introduced in the model do not compromise the reliability of the numerical results which correctly reproduce the \textit{de novo} anaerobic granulation process and are qualitatively in accordance with several experimental studies present in literature \cite{sekiguchi1999fluorescence,batstone2004influence,collins2005distribution}.

\subsection{Dimension, microbial distribution, ecology of biofilm granules and evolution of bulk liquid characteristics} 
\label{n5.2}

The dimensional evolution of the granule is described through the expansion of a spherical free boundary domain, whose radius varies over time. In particular, the \textit{de novo} granulation process is modelled starting from an initial condition where all the biomass present in the reactor is in planktonic form and there are no biofilm granules. Subsequently, the attachment process of the planktonic biomass initiates the granulation process and the granules begin to grow. The evolution of the granule over time is governed by the positive contributions of sessile metabolic growth and attachment flux of planktonic biomass and by the negative contribution of the erosive detachment flux. In the initial phase of the process, the size of the granule grows exponentially due to the high availability of substrates which leads to high rates of sessile metabolic growth, the high attachment flux induced by the presence of planktonic biomass within the bulk liquid and the negligible detachment flux, proportional to the granule size ($\sigma_d(t)=\lambda R^2(t)$). Later, the concentrations of soluble substrates and the planktonic biomass within the bulk liquid reduce. This leads to a reduction in metabolic growth and attachment flux. In addition, the detachment flux intensifies as a result of the granule size increase. Consequently, the growth of the granule decreases until it reaches a steady-state value regulated by the balance between the positive and negative mass fluxes.

The numerical studies show that the evolution of the granule dimension and its equilibrium value are deeply influenced by some factors, such as hydrodynamic shear stress, mass transfer of soluble substrates, composition of the influent and granulation properties of the planktonic biomass present in the inoculum. The model results presented in NS5 report that the granule dimension is governed by the shear forces: high shear forces produce an intense erosion on the granule surface and limit its growth. This qualitative result is fully in agreement with the literature \cite{liu2002essential,tay2001effects,arcand1994impact}. Since the shear forces depend on the hydrodynamic conditions occurring within the reactor, the model suggests that different particle distributions can be achieved by regulating the hydrodynamic regime in the reactor. As reported in NS2, the composition of the influent can affect the size of the granule due to the anabolic pathway of the species involved in the process. For example, the consumption of sugar induces a larger growth of sessile biomass compared to the consumption of the other substrates. As a result, for equal OLR, an influent with higher concentrations of sugar leads to the growth of more sessile biomass and larger dimensions of the granules. NS4 reports that larger amount of soluble substrates is consumed in denser granules, per unit of volume, therefore the metabolic growth rates decrease faster and lead to smaller equilibrium dimensions, in accordance with \cite{liu2002essential}. Finally, NS3 suggests that the granulation properties of the planktonic biomass regulate the evolution of the granule, especially in the initial phase of biofilm formation. More precisely, when the planktonic biomass is more prone to form biofilm structures and to grow in sessile form, the granulation process is more intense, and the granules grow faster.

In addition, the model describes the microbial ecology that develops in the granule and shows the distribution of the different microbial species. In the initial phase, the granulation process is governed by acidogenic and methanogenic species. This is in agreement with several studies \cite{trego2019granular,pol2004anaerobic,jian1993study} which report that methanogens play a fundamental role in the formation of the initial nucleus of the granule. Indeed, some methanogenic cells have a filamentous morphology \cite{trego2019granular} and ability to use quorum sensing strategies \cite{li2015significant,zhang2012acyl,li2014characterization} which improve their granulation properties and increase the efficiency of the granulation process. At the same time, acidogens have higher growth rates than other species, therefore they are present in large amount within the granule since the beginning of the process. Acidogens and methanogens exhibit the tendency to grow in different areas of the granule: the first in the outermost layer and the second in the inner part. This distribution becomes more evident over time. In particular, when the granule is mature, it is constituted by a large internal part populated by methanogens shielded by a thin external layer of acidogens. For long times, a homogeneous growth of acetogens and inert material deriving from the biomass decay is also observed. Many experimental studies \cite{sekiguchi1999fluorescence,batstone2004influence} show a microbial distribution within the anaerobic granules similar to the results proposed by the model. 

However, the model suggests that some factors such as the composition of the influent, the biomass density and the hydrodynamic shear forces developing in the reactor can radically affect the microbial distribution. For example, NS2 shows that in the case of higher concentration of sugar within the influent wastewater, the granules that develop during the process show larger amount of acidogenic biomass. Conversely, when only VFAs are present in the influent, acidogens do not develop and methanogens and acetogens dominate the granule. Furthermore, the model qualitatively reproduces the microbial distribution observed by \cite{batstone2004influence} in granules of different densities (study NS4). The biomass density deeply influences the mass transfer of the soluble substrates and therefore the microbial ecology. Accordingly, low densities lead to small gradients of soluble substrates across the granule and a homogeneous distribution of the microbial species is observed throughout the biofilm. More pronounced gradients develop within denser granules, leading to a more stratified distribution of biomass. According to NS5 study, shear stress conditions appear to be an additional factor affecting the microbial distribution in an anaerobic granule, playing an important role especially on the amount of active and inactive biomass: as mentioned above, intense shear forces result in the formation of small granules where there is high availability of substrates and the active biomass prevails over the inactive one; lower shear forces induces the formation of larger granules populated by high amount of inactive biomass which accumulates in the innermost zone of the granule due to lack or shortage of soluble substrates.

Furthermore, the model describes how the granulation process influences the characteristics of the bulk liquid and the effluent, in particular the concentration of planktonic biomasses and soluble substrates. The planktonic biomasses are present in the inoculum initially introduced in the reactor and represent the microbial community that initiates the granulation process. In the first few days, the concentration of planktonic biomass decreases rapidly as a consequence of attachment and dilution. The first concerns the aggregation of planktonic biomass which converts to sessile form and contribute to create granular structures. The second is the result of the hydrodynamic regime established in the reactor. As already reported, the HRT of granular-based systems is fixed small enough to guarantee hydrodynamic conditions and shear forces optimal for the granulation process. Such HRT values are highly unfavourable for planktonic biomass, which has short retention time to grow and consequently is diluted. After a few days, the planktonic biomass is completely washed out. Note that this result also derives from the assumption that biomass detached from the biofilm due to erosive phenomena does not constitute new planktonic biomass. Planktonic acidogens remain in the bulk liquid longer than other species due to their higher growth rate. Soluble substrates are produced or consumed in the bulk liquid due to the effect of planktonic and sessile biomass. In the initial phase, the granules are small and the substrates are converted mainly by planktonic biomass. Subsequently, when the size of the granules increases and the planktonic biomass has already been washed out, the trend of the substrates in the bulk liquid is governed by the granular biomass. Note that soluble substrates and planktonic biomasses achieve the steady-state values in a shorter time than sessile biomass within the granule. The effluent results to be purified at the steady-state, with the complete conversion of sugar and VFAs into methane.

Obviously, the trend of planktonic biomass and soluble substrates within the bulk liquid is influenced by the influent composition (study NS2). Furthermore, NS3 shows that the granulation properties of the planktonic microbial community initially present in the reactor deeply affect the velocity of the \textit{de novo} granulation process. When the granulation process occurs faster, the conversion rate of substrates is higher, and the process reaches the steady-state in a shorter time. Accordingly, the model results confirm that improving granulation properties by regulating hydrodynamic conditions, by adding substances which stimulate the quorum sensing \cite{li2015significant,zhang2012acyl} or by adopting bioaugmentation strategies \cite{nancharaiah2008bioaugmentation,jin2014bio} allows to speed up the process and reduce the long start-up times of anaerobic granular systems, which represent a critical issue in the operation of granular-based systems.


\section{Conclusions} 
\label{n6}

In this work, a mathematical model able to reproduce the \textit{de novo} granulation process involved in a generic granular biofilm system has been introduced. The work presents the derivation of the model equations which govern the expansion of the granule, the growth of sessile biomass, the transport of substrates and planktonic cells within the granule, under the assumption of radial symmetry. Such equations have been derived from mass balances considerations in the framework of continuum mechanics. Processes of growth and decay of attached and planktonic biomass, attachment from bulk liquid to biofilm, detachment from biofilm to bulk liquid, invasion of planktonic cells, conversion and diffusion of soluble substrates are modelled. The model has been applied to anaerobic granular systems to test its qualitative behaviour and study the process for a case of biological and engineering interest. The model takes into account the different contribution that individual microbial species can provide to the granulation process. For example, in the anaerobic case, the model is able to consider the fundamental role that some species of methanogens play in the granulation process by setting appropriate values of attachment velocity. The results shown describe exhaustively the anaerobic granulation process, the main properties of the granules such as the dimensional evolution, the ecology, the biomass distribution, the relative abundance and the evolution of the bulk liquid characteristics (soluble substrates and planktonic biomass). Finally, further numerical studies have been carried out to investigate the effects of some significant factors on the process.

The most interesting observations resulting from the numerical studies are reported below:

\begin{itemize}
 \item The anaerobic granule presents a typical microbial stratification: methanogens and acetogens populate the innermost layers of the granule and are shielded by a thin external layer of acidogens; the thickness of these layers depends on multiple factors (composition of the influent wastewater, biomass density, detachment forces).
\item Intense hydrodynamic conditions and short HRT typical of granular biofilm systems limit the growth of planktonic biomass, which is washed out.
\item The influent wastewater composition affects the evolution, ecology and microbial stratification of the granules.
\item The granulation properties of the planktonic biomass considerably influence the start-up period of the system. Strategies such as controlling hydrodynamic conditions, quorum sensing stimulation, bioaugmentation can reduce this period and enhance the start-up efficiency.
\item The density of biomass regulates the mass transfer of soluble substrates and consequently the microbial distribution within the granule: denser granules have a more layered structure while less dense granules have a more homogeneous structure.
\item Shear forces have a large impact on the granule size and the ratio of active to inactive biomass: higher shear forces lead to smaller granules constituted by larger fractions of active biomass.
\end{itemize}

The results shown are qualitatively in accordance with the experimental evidence from the literature. Accordingly, this model is able to correctly simulate both the formation and maturation of anaerobic granules by focusing on both the transient and the steady-state. From an engineering point of view, this leads to conclude that the model proves to be a useful tool both in the start-up and in the purification process of anaerobic granular biofilm systems. Furthermore, the model can be applied to any biological process which takes place in a granular-based system by choosing the suitable model variables and kinetic expressions.

In any case, some model parameters such as the values of the attachment velocities of the planktonic biomass are introduced here for the first time and should be calibrated and validated on the basis of experimental data. Finally, in a future work, the detachment process leading to the breaking of the granule and the consequent formation of new granules could be included in the model, in order to describe the entire life cycle of biofilm granules.
 
 \section*{Acknowledgements}

Alberto Tenore's research activity was conducted in the context of D.D. n. 1377 on June 5, 2017, additional PhD fellowships for 2017/2018 academic year, course XXXIII within the framework of the "Programma Operativo Nazionale Ricerca e Innovazione (PON RI 2014/2020) Action I.1 - Innovative PhDs with industrial characterization".

Fabiana Russo's research activity was conducted in the context of D.D. n. 155 on 17 May 2018 additional PhD fellowships for 2018/2019 academic year, course XXXIV within the framework of POR Campania FSE 2014-2020 ASSE III - Specific objective 14 Action 10.5.2 - Public notice "Innovative PhD with industrial characterization".

The authors also acknowledge the support from: CARIPLO Foundation (progetto VOLAC, Grant number: 2017-0977); Progetto Giovani G.N.F.M. 2019 "Modellazione ed analisi di sistemi microbici complessi: applicazione ai biofilm".

This paper has been performed under the auspices of the G.N.F.M. of I.N.d.A.M.  

\bibliographystyle{unsrt}      

\bibliography{Tenore_et_al}  

\end{document}